\documentstyle[12pt]{article}
\begin{document}

\def\ir{I\kern-.1667em R\ }
 
\def\cd{\cdot
}\def\nl{\hfill\break}       
\def\np{\vfill\eject}       
\def\ns{\vskip 2pc\noindent}       
\def\nsect{\vskip 2pc\noindent}       
\def\nli{\hfill\break\noindent}       
\def\ni{\noindent}
\def\IR{I\kern-.255em R}
\def\Dl{\Delta}
\def\Del{\Delta}
\def\dpp{\delta p}
\def\dI{\delta I}
\def\xb{{x} }
\def\pb{\overline{p} }
\def\tb{\overline{t} }
\def\vb{\overline{v} }
\def\vzx{v_z(x)}
\def\mux{\mu(x)}
\def\eps{\epsilon}
\def\epsL{\epsilon_L}
\def\epsv{\epsilon_v}
\def\vbL{{{\overline{v} \over L }}}
\def\vbLd{{{}}}
\def\La{{{L \over a }}}
\def\aL{{{a \over L }}}
\def\om{\omega}
\def\xe{{x \over \eps}}
\def\xmn{x_{m,n}}
\def\zmn{z_{m,n}}
\def\xr{x_{m,n}}
\def\omm{\omega_{m,n}}
\def\omn{\omega_{m,n}}
\def\omt{\tilde{\omega}}
\def\umn{\mu_{m,n}}
\def\Tt{\tilde{T}_{m,n}}
\def\Ttw{\tilde{T}_{}}
\def\Ttl{\tilde{T}_{}}
\def\etab{\overline{\eta} }
\def\rlrm{({r_l \over r_m}) }
\def\rlw{({r_l \over w}) }
\def\twp{2 \pi}
\def\dxm{{ {x} - {x}_{mn} \over \eps_L \Dl_m} }
\def\vvt{\tilde{\vec{v}}}
\def\kv{\vec{k}}
\def\vv{\vec{v}}
\def\phiv{\vec{\phi}}
\def\vt{\tilde{{v}}}
\def\xt{\tilde{{x}}}
\def\Kt{\tilde{{K}}}
\def\xv{\vec{x}}
\def\xh{\hat{x}}
\def\yh{\hat{y}}
\def\zh{\hat{z}}
\def\thh{\hat{\theta}}
\def\Kb{\overline{K} }
\def\Kh{\hat{K}}
\def\Kbh{\hat{\overline{K} }}
\def\Dbh{\hat{\overline{D} }}
\def\Th{\hat{T}}
\def\The{\hat{T}_{\eps}}
\def\Wh{\hat{W}}
\def\Yh{\hat{Y}}
\def\kpb{\overline{\kappa}}
\def\Dbh{\overline{D}_h}
\def\Dh{\hat{D}}
\def\Gh{\hat{G}}
\def\vd{\vec{v} \cdot \nabla}
\def\ckp{{\epsv^2 \vb^2 \over L}}  
\def\ckpd{{\epsv^2 \over \delta^2 }}  
\def\ckpl{{\epsv^2 L^2\over a^2}}  
\def\ckplds{{\epsv^2 L^2\over \delta a^2}}  
\def\sst{\scriptstyle}
\def\ssst{\scriptscriptstyle}
\def\cl{\centerline}
\begin{center}
{\bf QUASILINEAR DRIVEN TRANSPORT IN A SHEARED FLOW FIELD\\}
{\bf K.S. Riedel \\
Courant Institute of Mathematical Sciences \\
New York University \\
New York, New York 10012}
\end{center}
\begin{abstract}
The evolution of a passive scalar field
is considered for a slowly varying
stratified medium, which is convected in an incompressible
sheared flow with many
overlapping static flux islands.
Within the quasilinear/random phase approximation, a multiple scale expansion
is made. Due to the rapid spatial variation of the temperature, the
``ensemble'' averaged/ slowly varying part of the solution is not described
by the arithmetic average of the  oscillatory evolution equation.
The standard Markovian and continuum approximations are shown to be
invalid.
For times of order $N$, where there are $O(N^2)$ excited modes, 
most of the time dependent perturbation
phase mixes away and
the fluid reaches a new saturated state
with small time oscillations about the temperature.
This saturated state has smaller resonance layers, 
(corresponding to magnetic islands) than 
those that occur in the isolated 
resonant perturbation case. Thus the quasilinear 
response to the resonant interactions
reduces the effective size of the perturbations.
The temperature gradient of the saturated state vanishes at all
the excited resonance surfaces but has a nonzero average.  
Thus either the quasilinear approximation ceases to be valid on long
time scales, or
the fluid remains essentially in this modified equilibrium and does not  
evolve diffusively.
Thus collisionless, driftless fast particles will not be lost 
rapidly in equilibria with many small islands.  
\end{abstract}

\np
\noindent
{\bf I. INTRODUCTION}

The passive scalar evolution equation
is widely studied as a model problem for turbulent systems.
One class of research concentrates on self-consistent approximate
solutions of the fully nonlinear convective nonlinearity under
the simplifying assumptions of homogeneity and isotropicity$^1$.
These assumptions eliminate much of the geometry of the underlying
stationary state. We consider systems where a weak ensemble of
turbulent modes$^{2,3}$ is superimposed on a strongly sheared equilibrium
flow. The response of the passive scalar field is resonant at  
the points where the turbulent wavevector, $\kv$, is perpendicular 
to the equilibrium flow velocity.

When the fluctuations are small,
three wave interactions are a higher order effect and may be neglected
in the evolution of the turbulent modes. Instead the turbulent
modes interact only by modifying the equilibrium gradients. Thus
the turbulent modes
evolve linearly on a slowly varying equilibrium state. 
The turbulent modes only affect the evolution of the background
state.
In the turbulence literature, this is known as the  
``weak coupling approximation",
the  ``quasilinear approximation",
or the ``random phase approximation." 
In the standard weak turbulence theory of Galeev and Sagdeev$^3$,
the weak coupling truncation is the first of four formal steps.
The last three formal steps, the Markovian and continuum ``approximations"
and the spatial averaging of the coefficients,
will be examined using a multiple scale expansion.
{Since our analysis is based exclusively on the quasilinear/random
phase approximation, our conclusions take the form: ``If the truncated
system of equations remains valid for long times, then ...."}

We consider the driven passive scalar transport problem in a
sheared torus.
We note that the passive scalar evolution problem is isomorphic
to the Liouville equation for the
diffusion of trajectories in a nearly integrable Hamiltonian
system with one and a half degrees of freedom$^{2}$. The problem also is
equivalent to field line diffusion in magnetohydrodynamic equilibria
with magnetic islands. Past analysis of the transport of particles
in stochastic field lines has been based on the quasilinear diffusion
theory for field line diffusion$^{4,5}$.
Our work is motivated by the pioneering analysis
of Rosenbluth, Sagdeev, Taylor and Zaslavskii$^{2}$.
    
We consider the effective diffusion of the passive convective scalar
equation in a torus:
$$
{ \partial {T}(\xv,t)   \over  \partial t}
= -\vv(\xv,t)_{} \cdot  \nabla  T(\xv,t)  
\eqno(1)
$$
where $\vv(\xv,t)$ is a given velocity field.
We use coordinates $x,\theta,z$, where $x$ is the stratified direction,
$\theta$ is the poloidal direction, and $z$ is the toroidal direction.
The torus has a toroidal radius of size L and a
poloidal radius of size $a$. We assume the aspect ratio, $\La$, is order one.

Both the driven velocity field, $\vv(\xv,t)$, and the passive scalar,
$T(\xv,t)$, are $\twp L$ and $\twp a$ periodic in the toroidal and
poloidal planes. Toroidal periodicity is the natural boundary 
condition for the breakup of integrable surfaces in Hamiltonian
systems and for toroidal magnetic confinement systems.
Toroidal periodicity quantizes
the toroidal mode spectrum. 
In contrast, fluid
flow in an infinite  pipe
has a continuous spectrum of resonant modes.

We assume that the equilibrium velocity field is
sheared:
$\vv_o(x) = \vb [\vzx \zh - \aL \mux \thh ]$
and the equilibrium temperature field is stratified in the $x$ direction:
$ {T_o}(\xv,t)= {T_o}(x,t)$. 

Since the temperature is purely convected,
the natural initial boundary value problem is that the temperature is 
given both on the boundary and in the interior at $t =0$ and the boundary
values are convected as well. If the normal velocity at the boundary vanishes,
our fixed boundary problem is well posed. If in addition, 
we assume the temperature
at the boundary initially depends only on the $x$ variable, the
temperature will remain constant on the boundary. If the normal velocity at
the boundary does not vanish, then the problem is wellposed as a
free boundary problem. 

When the normal velocity at
the boundary vanishes, the conservation formulation guarantees that there
is no net flux of temperature ot of the domain and that temperature
is conserved pointwise in Lagrangian coordinates. As the number and strength
of the helical perturbations increases, the trajectories of
neighboring temperature surfaces will become increasingly distorted.
The $\theta-z$ surface average of these distorted, oscillating temperature
survaces may then tend to flatten and approximate diffusion. In
this article, we examine whether this often proposed mixing/Hamiltonian
diffusion occurs in our model toroidal problem.

In the $x-\theta$ plane, the characteristic length is $a$, and in the toroidal 
direction, the characteristic length is $L$. 
The characteristic velocity of the equilibrium flow is $\vb$. Our time
is measured in units of $L / \vb$. 

We consider the evolution of the stratified equilibrium in an
ensemble of many small, helical perturbations.
The velocity field is incompressible, $\nabla \cdot \vv(\xv,t) = 0$. 
We decompose the perturbing velocity field, $\vvt$, into a sum of
discrete Fourier harmonics:

$$\vvt(\vec{x},t)=
\epsv \vb \sum_{m,n \neq (0,0)}^{}
\vv_{m,n}(x) e^{i( m{\theta\over a} + n{z\over L} -\omm \vbL t)}
\eqno(2)
$$
where $\vvt_{-m,-n}= \vvt_{m,n}  $and $\om_{-m,-n}= \om_{m,n}$. 
The $\omm$ are the eigenfrequencies of the excited modes 
and are assumed to be real. 
If we were interested in infinitely long pipes, the sum over
$n$ would be replaced by an integral $dn$. We shall show that the
often made substitution of $\sum_n \ \rightarrow \ \int dn$ is not valid in
our system. Thus the quantization of the mode spectrum induced by the
boundary conditions is an essential aspect of our problem.

The mode amplitudes, $\vt_{m,n}$, may be treated as fixed,
or as random variables.
Since the $\vt_{m,n}$ do not have any time dependence,
the correlation time for the mode amplitudes $|\vt_{m,n}|^2$ is infinite.  
We will show that static or strictly oscillatory perturbations 
are pathological and do not generate diffusion to second order in the
standard amplitude expansion.  
We note that the finite autocorrelation case may behave diffusively
and that our results are limited to velocity perturbations with
an infinite mode autocorrelation time. 
{In Ref. 5, Krommes refers to this same problem as the short/zero 
autocorrelation case. Although the autocorrelation function, 
$<\vt,\vt>$, has an infinite autocorrelation time, 
the effective $<\vt,\Ttw>$ correlation decays initially
with velocity shear time, $1/\vb\partial_x \mu(x)$,
in both the Eulerian and Lagrangian coordinates. However for
long times, the Eulerian autocorrelation function of
$<\vt,\vt>$ ceases to decay and therefore has an infinite autocorrelation
function and Kubo number. In Appendix B, we show that the effective Lagrangian 
autocorrelation function decays on the shear time scale until it saturates
at a small constant level. 
Furthermore, as pointed out in 
Ref. 5, the neglect of the fully nonlinear terms in this analysis is 
equivalent to assuming the Kolmogorov exponentiation time or
nonlinear decorrelation time is effectively infinite with respect to
autocorrelation time.}

We suppress the dependence of $\vt_{m,n}$ on the random variable,
$\omega$, and in Appendix F,
consider the additional smoothness due to random
Doppler shifts. Our results include the case where the eigenfrequencies, $\omega_{m,n}$,
are also random but time independent.
 
We scale the fluctuation
amplitudes with the small parameter, $\epsv$. 
We scale the mode spectrum with a second small parameter, $\eps_L$.
If we did not scale the mode spectrum, then for small
enough $\eps_v$, closed flux surfaces would exist and there would be no 
net diffusion. Thus as $\eps_v$ decreases, we must increase the number of
excited modes to ensure all flux surfaces are broken.

Our scaling of the mode spectrum
must not only guarantee that the Chirikov overlap criteria 
is satisfied everywhere, but also that the quasilinear truncation is
formally valid. Within these restrictions, we believe our multiple 
scale expansion of the long time quasilinear evolution will be valid  
for a large class of mode spectra. (See Appendix A.) 
We assume a strict cutoff in mode amplitudes
at the values $N_m$ and $N_n: |\vvt_{mn}|\equiv 0$ if $|m|>N_m$ or $|n|>N_n$.  
We scale the mode cutoffs as $N_n \sim 1/\eps_L$ and 
$N_m \sim 1/\eps_L^{\alpha}$ where $0 \le \alpha \le 1$. $\eps_L$ may also
be thought of as the characteristic scalelength for nonresonant
phenomena. For
fixed poloidal mode number, $m$, the distance between adjacent
resonance surfaces is proportional to $\eps_L$.
On the short wavelength
scale, $y \equiv \epsL x$, the number of excited rational
frequencies is proportional to $N_m$.  
For convenience, we assume that all of the excited modes have
roughly similar amplitudes.


The fluctuating temperature perturbation is 
decomposed into Fourier harmonics, $\Tt$, which evolve
temporally as

$${
{ \partial {\Tt}(x,t)   \over  \partial t} + i \vbL (n\vzx -m\mux-\omm){\Tt}(x,t)    
\ = } $$ $$  
 - {\epsv \over a}\vt_{m,n}   (x){\partial  T_o\over \partial \xb}(x,t)   
- {\epsv} \left( \vvt(\xv,t)_{} \cdot  \nabla  \Ttw(\xv,t) \right)_{m,n} \ ,
\eqno(3)
$$
where we have used the nondimensional $x$ variable, $x \equiv x_{dim} / a$.
The temperature perturbations, $\Tt$, are spatially localized in small 
neighborhoods where $(n\vzx -m\mux-\omm)$ is not large and therefore $\Tt$ 
have a spatial localization of size $\eps_L$.
Due to the spatial localization of $\Tt$,
the nonlinear convective term,
$ \left( \vvt(\xv,t)_{} \cdot  \nabla  \Ttw(\xv,t) \right)_{m,n} $,
is the sum of $N_n$ terms of roughly order $\epsv^2 / \epsL$.  When the orientation of the phases is random
with respect to each other, the sum scales as $N_n^{1/2}$ times the size
of an individual element.  The random phase approximation postulates that
the sum of $N_n$ terms probabilistically has a zero mean value and
thus only its variance determines the magnitude of the nonlinear term.
Thus, in the random phase approximation, the nonlinear convective term,
$ \left( \vvt(\xv,t)_{} \cdot  \nabla  \Ttw(\xv,t) \right)_{m,n} $,
is order $N_m^{1/2} \epsv / \epsL$ smaller
than the $ \vt_{m,n}   (x){\partial  T_o(x,t) \over \partial \xb}$.
For the quasilinear mode truncation to be valid on the ideal timescale,
{\it we assume that $\epsv << \epsL N_m^{1/2} $ }, in addition to the
random phase ansatz.
Thus the eddy turnover time, $\epsL/\eps_v$, is assumed to be large.

  
For long times, the truncated equations may cease to reflect the actual
physical system. Nevertheless, the truncated equations are interesting and
widely used. Furthermore, the truncated system is analytically tractable
using a multiple scale expansion.
We concentrate on a detailed examination of the 
properties of the quasilinear equations.  
The quasilinear or ``weak coupling" equations for the Fourier harmonics are:

$$
{ \partial {\Tt}(x,t)   \over  \partial t} + i \vbL \umn(x){\Tt}(x,t)    
= - {\epsv \over a}
\vt_{m,n}(x){\partial  T_o\over \partial \xb}(x,t)   \ ,
\eqno(4)
$$
where we have defined $\umn(x)$ to be the resonance denominator,
$\umn(x) \equiv m \mu(x) - n\vzx - \omn$. We denote by $\xmn$, the 
location of the spatial resonance where $\umn(x)$ vanishes.
This first order differential equation in time may be integrated to
yield

$$
 {\Tt}(x,t)=  - {\epsv \over a} \vb \vt_{m,n}(x) \int_{s=0}^{t}
 e^{i \vbL \umn(x)(s-t)}
 {\partial  T_o(x,s) \over \partial x} ds  \ .
\eqno(5)
$$

In contrast to dissipative systems, {\it this ideal system possesses
infinite memory as indicated by the lack of a temporal decay
constant in the kernel of Eq.(5). } At the resonance surface, $\xmn$,
$\Tt$ grows linearly in time unless $\partial_x T_o$ vanishes at
the resonance surface.
We rescale time, 
$\tb \equiv \delta t$, where the small parameter $\delta$ is a
yet unspecified function of $\epsv$ and $\epsL$. 
If $T_o(x,t)$ were to vary only on the slow, $1/\delta$ timescale,
the magnitude of $\Tt$ can be estimated by replacing
$T_o(x,s)$ by  $T_o(x,t)$ in the integrand. We find that
 $\Tt \sim O( {\eps_v } A_{m,n}(x) 
{  sin ({ \vbLd \umn(x) {\tb\over \delta} } )
\over  \vbLd \umn(x) } )$. Within $O(\eps_L \delta)$
of the resonance surface, $\xmn$, $\Tt$ would be order
${\eps_v \over \delta}$. This linear in time growth of
$\Tt$ at the resonance surface could easily violate
the quasilinear ordering.
However, we will show that
the growth of the $\Tt$ saturates due to the flattening of 
$\partial_x T_o$ at the resonance surfaces.

The quasilinear evolution equation for the $m=0,n=0$ mode is

$${
{ \partial {T_o}(x,\tb)   \over  \partial \tb}
= {\epsv L\over \delta a} {\partial  \over \partial \xb}  
\sum_{m,n \neq (0,0)}^{}
\vt_{-m,-n}(x,\tb) \Tt(x,\tb) =
 }$$ $$
\ckpd {\partial  \over \partial \xb}  
\sum_{m,n \neq (0,0)}^{}
A_{m,n}(x) \int_{s=0}^{\tb}
 cos ({ \vbLd \umn(x){(s-\tb)\over \delta}} )
{\partial  T_o \over \partial \xb}(x,s)  ds  ,
\eqno(6)
$$
where $A_{m,n}(x) \equiv (L/a)^2|\vt_{m,n}(x)|^2$ and the nondimensional $slow$
time, $\tb \equiv {\delta vt_{dim} /L}$ has been used. 
{\it We concentrate on analyzing the
 mathematical properties of Eq. (6) in the limit
of many small, short wavelength perturbations.} 

In the next section, we summarize the pioneering analyses of
Rosenbluth, Sagdeev, Taylor and Zaslavskii$^2$ and the related
work by Krommes, et al.$^5$. We derive a number of mathematical
results for Eq. (6) and its Laplace transform, Eq. (12) in Sec. III.
We show that Eq. (6) has purely continuous spectrum, except for an eigenmode
at zero frequency. 

Sec. IV reviews the homogenization theory of rapidly varying
dielectric functions.
In Sec. V and Appendix E, we make  multiple scale expansions to determine
asymptotic behavior of $\Th(x,q)$. 
Sec. V addresses the overlapping resonance 
case and Appendix E addresses the isolated 
resonance layer case.
In Sec. VI, we determine the time evolution of $T(x,t)$ by 
inverting the Laplace transform. We show that the truncated
system phase mixes to a new Cantor-set like gradient on a time
scale of order $N_m$.  We conclude by discussing our results
for the quasilinear system and their relevance for the actual
physical system.

In Appendix A, we discuss the physical basis of our rescaling of the
mode spectrum.
In Appendix B,
we make an equivalent second order truncation in Lagrangian 
coordinates and show that an equivalent time evolution occurs.
In Appendix C, we examine the behavior of the coefficients for
small values of the Laplace parameter, $q$, and show that
Eq. (12) is analytic in a neighborhood of order $1/N_m$ about
the isolated pole  at $q=0$. 
In Appendix D,
we examine the behavior of $\Th(x,q)$ at the real and removable
singularities. 
In Appendix F, we discuss the modifications which result if
the eigenmode frequencies, $\omm$ are random, but time independent.  

\np
{\bf II. SUMMARY OF PREVIOUS FORMAL ANALYSES}
 
This same quasilinear equation, Eq. (6), was originally derived and analyzed
by Rosenbluth et al. in Ref. 2. Rosenbluth et al. found that for
very small perturbations, a new modified equilibrium state is 
reached with the new equilibrium temperature gradient vanishing
at each excited rational surface. As the perturbations increase, 
the resonance layers overlap. In Ref. 2, an effective diffusion
equation was derived for this ``overlapping resonance" regime.
We rigorously rederive and systematize the results of  
Ref. 2 for the ``isolated resonance"  regime.  
In contrast, in the ``overlapping resonance" regime, we  continue
to find nearby saturated temperature states which do not decay
on a slow diffusive timescale.

We briefly review the standard derivation$^{2}$ of an effective
diffusion equation for Eq. (6), emphasizing the differences
between the traditional approach and a careful multiple scale
analysis. We note that no part of the following arguments 
of Ref. 2 uses the Chirikov criteria of resonant layer overlap.
 
Since the evolution of the slowly varying part
of $T_o(x,t)$ is of primary interest, previous calculations 
ignored the rapidly varying part of $T_o(x,t)$, replaced $T_o(x,s)$ by  
$T_o(x,t)$, and performed the temporal integration. 
This formal step is called the ``Markovian approximation"
in the turbulence literature and is one of the principal objects
of our investigation. 
The resulting
heat equation has a time dependent heat conductivity:

$$ \kappa(x,\tb) =
\sum_{m,n \neq (0,0)}^{}
\ckplds A_{m,n}(x) 
{  sin ({ \vbLd \umn(x) {\tb\over \delta} } )
\over  \vbLd \umn(x) }  \ .
\eqno (7)$$
In general, {\it this approximation
is invalid because the rapidly varying kernel will induce rapid 
variations in $T_o$.} Although these variations in $T_o$ are small,
the variations in the gradients of $T_o$ are order one and 
therefore influence the time evolution of $T_o(x,t)$.

The second step in the traditional
analysis is to approximate the sum over toroidal wave number by
an integral over $n$ space. This ``continuum" approximation
is only valid when the integrand is a slowly varying function of 
 $n$. Unfortunately, {\it the integrand is a rapidly varying function of 
 $n$.}

Naively, one hopes that the ``Markovian approximation" and
the ``continuum approximation" would become valid in the limit of 
large shear. If the shear were to be scaled such that it tended
to infinity as the modes strengths tended to zero, the density of
mode rational surfaces would become large and the kernel term,
$cos ({ \vbLd \umn(x) {\tb/\delta} } )$ would be rapidly varying 
away from the resonance surfaces, $x_{mn}$. However the
``Markovian approximation" and the ``continuum approximation"
are questionable even in the large shear limit, because
the selfconsistent solution, $T_o(x,t)$, will also vary 
rapidly as the shear tends to infinity.


The third step in the standard approach is to 
replace $\kappa(x,t)$ by its limiting form in the long
time and spatially averaged limit. 
In Ref. 2, this step is divided into two parts. First,
Rosenbluth et al. state that
$sin( \alpha t) / \alpha \rightarrow \pi \delta(\alpha)$ as
$  t \rightarrow \infty $. 
Implicitly, Rosenbluth et al. are assuming that the spectrum,
$A_{m,n}$ depends only weakly on $n$.
We do not understand this third formal step of Ref. 2 if the
toroidal harmonics are different. Therefore we restrict our
to $A_{m,n} \equiv A_m + o(1)$ and $v_z(x) \equiv 1$
in the rest of this paragraph.
Thus Rosenbluth et al. replace Eq. (7) with
$$\kappa_{lim} =    
{\pi \epsv^2 \over  \delta } ({L \over a})^2 
\sum_{m \neq 0}^{} {A_{mn}(x_{mn})\over \mu'} \delta(x-x_{mn}) \ .
\eqno (8a)
$$ 
Finally,
Rosenbluth et al. spatially average their expression for $\kappa(x)$:

$$\kappa_{lim} =    
{\pi \epsv^2 \over \delta \mu' } ({L \over a})^2 
\sum_{m \neq 0}^{} {A_{m}(x_{mn})} \ .
\eqno (8b)
$$ 
We believe that 
this complicated third step is trying to replace the rapidly
varying $\kappa(x,t)$  by a smooth limiting function.
However, $\kappa(x,t)$ 
tends only weakly, not pointwise, to $\kappa_{lim}$. 
The mathematical theory of homogenization$^{6-8}$ has shown
that the {\it limiting solutions to rapidly varying heat 
equations do not, in general, converge to the solution
corresponding to the limiting heat conductivity
$\kappa_{lim}$}.  

Krommes et al.$^5$ have reanalyzed a similar problem: the motion
of particles on chaotic field lines with a small random walk superimposed.
We note that the renormalization procedure proposed by Krommes has
never been performed in any detailed or explicit fashion and thus 
unexpected difficulties may occur. Furthermore, the renormalization
theory is purely formal and makes many of the same formal manipulations
as the quasilinear theory. Thus the difficulties
of the quasilinear theory, which we examine in this article, 
need to be addressed in renormalization theories as well.

Krommes replaces the continuum approximation for toroidal mode 
numbers with a continuum approximation for $k_{\parallel}$,
because he considers that the kernel varies more slowly
with respect to $k_{\parallel}$ than $k_z$. In reality, the
kernel has order  one dependencies on $k_{\parallel}$ for
finite time and fast variation with respect to $k_{\parallel}$
for long times. The $k_{\parallel}$ continuum approximation has the
other difficulties that $k_{\parallel}$ is a spatially varying
function and that the Fourier modes are unlikely to be 
averagable in the $k_{\parallel}- k_{\theta}$ plane. In other
words the density of modes is nonuniform in the
$k_{\parallel}- k_{\theta}$ plane and there is no symmetry 
such as the ``equivalence of harmonics"$^{9-12}$.

A number of articles study the $1 \ 1/2$ degree of freedom analog
of the passive scalar equation and report to derive 
quasilinear diffusion coefficients$^{13-14}$. 
These authors consider a single set of resonances in the 
limit that the individual mode amplitudes tend to infinity
($\eps_v \rightarrow \infty$).
Our limit, many small amplitude
modes interacting with each other by modifying the underlying
equilibrium seems to not only be the more physically relevant
limit, but also the limit which is more appropriately termed
quasilinear. 
This $\eps_v \rightarrow \infty$ limit 
corresponds to scaling the velocity
shear to infinity as $\partial_x \vv_o(x) \sim 1/\eps_v^2$!
In the `derivation' of
Rechester and White for this infinitely large shear limit, they
replace rapidly oscillating coefficients with their weak limit.
As in Ref. 2, the solution of the limiting equation need not be
the limit of the solutions to the actual equations. 
 
\newpage
\nsect
{\bf III. PROPERTIES OF THE QUASILINEAR EQUATION}

In this section, we derive some fundamental mathematical
properties of Eq. (6); in particular, that the spectrum of Eq. (6) is
purely continuous and oscillatory except for the steady state eigenmode. 
Equation (6) is
a convolution differential equation of the form:  

$$
{ \partial {T} \over  \partial \tb}(x,\tb) =
 { \partial \over  \partial x} \int_0^{\tb} 
K(x,{{\tb - s}\over \delta})
{ \partial {T}\over  \partial x}(x,s) ds   ,
\eqno(9)
$$ 
where we have suppressed the subscript, o.
The time evolution of the mean squared profile satisfies
$$ {\partial\over\partial t} 
\int_{-\infty}^{\infty}{{T}^2}(x,\tb) =  
- \int_{o}^{\tb} \int_{-\infty}^{\infty}
{ \partial {T}\over  \partial x}(x,\tb) 
 K(x,{(\tb-s)\over \delta})
{ \partial {T}\over  \partial x}(x,s) ds    \ .
\eqno(10)
$$
Since $ K(x,{(\tb-s)\over \delta})$ is positive at the resonance
surfaces  
and is rapidly oscillating elsewhere, Eq. (10) suggests that
$T(x,\tb)$ might phase mix and decay. 
However, Eq. (10) also allows predominantly static solutions
where the gradient of $T$ vanishes at all resonances.
Furthermore, this simple analysis neglects
the rapidly oscillating part of  $T(x,\tb)$. 

Provided that ${ \partial {T}\over  \partial \tb}(x,\tb)   $ is
exponentially bounded, this convolution equation may be simplified
using the Laplace transform,

$$ \Th(\xb,q)= 
i\int_{o}^{\infty} e^{-iq\tb} T(\xb,\tb) dt , \ \ \ 
T(\xb,\tb)= {1 \over \twp i}
\int_{-ic- \infty} ^{-ic+\infty} e^{iq \tb} \Th(\xb,q) dq \ .
\eqno(11)
$$
For convenience, we have rotated the Laplace variable plane by making
the substitution, $p \equiv iq$, and taken $\Th(x,q) \equiv i\Th(x,p)$,
where $p$ is the normal Laplace transform variable.
This rotates the contour of
integration to 
an integration below the real $q$ axis.
Noting that $dp \over p$ is transformed to $dq \over q$,
the equation for $\Th(x,q)$ becomes:

$$
{ \partial \over  \partial \xb} \Dh(\xb,q)
{ \partial {\Th}\over  \partial \xb}(\xb,q)    - {\Th}(\xb,q)
= -{T}(\xb,t=0) /q  ,
\eqno(12)
$$
$$where \ \ \  \Dh(x,q) =
\ckpl
\sum_{m,n \neq (0,0)}^{}
{  A_{m,n}(\xb) \over \left(  
  { \vbLd \umn(x)}^2 - \delta^2 q^2 \right) },
\eqno(13) $$ and  $\Kh(x,q) = q \Dh(x,q)$ .
The kernel, $K(x,t)$, of Eq. (6) has the special property
that $\Kh(x,q)$ vanishes everywhere at $q=0$. Therefore the
substitution, $\Kh(x,q) = q \Dh(x,q)$, is justified. An
immediate consequence is that $\Th(x,q)$ has a pole at $q=0$
for all values of $x$. Since we will later show that this is
an isolated singularity, the pole at $q=0$ generates  the 
time independent part of the solution.
 
Equation (12) possesses the following symmetries:
$\Th(x,-q) = -\Th(x,q)$ and
$\Th(x,q^*) = \Th^*(x,q)$.
The second symmetry guarantees the reality of 
$T(x,t)$.
We define the Doppler shifted resonances, $\xmn^{\pm}(q)$, 
by $\umn(\xmn^{\pm}(q))= \pm \delta q$. 
Note that $\xmn^{\pm}(q=0)\equiv \xmn$.

We examine wellposedness and solution properties of the quasilinear
Eq. (12). Representations of the solutions are given in Sec. V and Appendix
E. We begin by
studying the solution for large values of the Laplace parameter $q$.
Expanding Eq. (12) in powers of $1/q$, we find that
$q {\Th}(x,q)- { {T}(x,t=0) }= O(1/q^2)$ and therefore ${\partial T
\over \partial t}$ exists and is exponentially bounded.


The poles of $\Dh(x,q)$ generate
removable singularities of Eq. (12).
To show this, we make the substitution, 
$\Yh(x,q) \equiv \Dh(x,q) { \partial \Th\over  \partial x} (x,q) $
which yields

$$
{ \partial^2 {\Yh}\over  \partial x^2}(x,q)    -  \Dh(x,q)^{-1}{\Yh}(x,q)
= {1 \over q}  { \partial {T}\over  \partial x} (x,t=0)   \ .
\eqno(14)
$$
Thus ${ \partial \Th\over  \partial x}(x,q)  $
vanishes at the poles of $\Dh(x,q)$.
Away from the zeros of $\Kh(x,q)$, Eq. (12) is clearly wellposed and depends
analytically on the Laplace parameter $q$. 
At the double poles at $q=0$, $\Th(x,q)$ remains analytic in $q$ and
both $\partial_x \Th(x,q)$ and $\partial_x^2 \Th(x,q)$ vanish at $x_{mn}(q=0)$. 
Thus $\Dh(x,q)^{-1}$ plays the role of a quantum mechanical potential,
$V(x,q)\equiv \Dh(x,q)^{-1}$.

Both Eq. (12) and Eq. (14) are formally selfadjoint equations of the
form $ (a_o(x)u')' - a_2(x)u =f(x)$. 
Therefore we use $u(x)$ to denote either $\Th(x,q)$
or $\Yh(x,q)$.
To construct the Greens function, we let $u_L(x,q)$ 
be a solution of $ (a_o(x)u')' - a_2(x)u = 0$, 
which satisfies the left boundary condition,
and $u_R(x,q)$ satisfy the right boundary condition.
The Greens function representation is:

$$ A(q)u(x,q) = u_R(x,q) \int_{a} ^{x} u_L(\xi,q) f(\xi,q) d\xi 
+ u_L(x,q) \int_{x} ^{b} u_R(\xi,q) f(\xi,q) d\xi  \ ,
\eqno (15)$$
where $A(q) \equiv a_o(x,q) [u_L'(x,q)u_R(x,q)-u_L(x,q)u_R'(x,q)] $ 
is independent of $x$.

 To examine the existence of normal modes, we multiply
Eqs. (12) and (14) by 
${ \Th^*}(x,q) $
and
${ \Yh^*}(x,q) $ respectively
and integrate by parts. 
The imaginary part of $\Dh(x,q)$
never vanishes except on the real and imaginary $q$ axes.  
Thus the imaginary part of the variational principle prevents the
existence of normal modes.
Clearly no normal modes occur when
$Re(q)=0$ and $q \ne 0$. Thus normal modes can occur only on the real
$q$ axis. 

For $Im(q)=0$, the energy in the variational principle
is unbounded unless the normal mode 
is analytic at each resonance surface. For two or more resonance
surfaces, this behavior is nongeneric.
The same arguments apply to the normal modes of Eq. (14).

The contour of integration in the inverse Laplace transform
can clearly be deformed up to the real $q$ axis
and be replaced
by a contour around the continuous spectrum on the real $q$ axis.
The symmetries of $\Th(x,q)$ imply that Eq. (11) can be rewritten as
$$
T(\xb,\tb)= {4 \over \twp } lim_{\eps \rightarrow0}
\int_{0} ^{\infty} Imag(\Th(\xb,q_R- i\eps) cos{q_R\tb}  dq_R \ .
\eqno(16)
$$

\bigskip
{\bf IV. HOMOGENIZATION OF RAPIDLY VARYING EQUATIONS }

Homogenization theory$^{6-8}$ considers an analogous problem of
estimating the effective heat conductivity of a composite material
with a rapidly varying thermal conductivity. Due to the excitation
of rapidly varying, order one gradients, the homogenized thermal
conductivity is the harmonic and not the arithmetic mean.
A more precise statement of the basic theorem of homogenization
follows. Consider the sequence of elliptic problems,

$$
{ \partial \over  \partial x} \kappa(\xe)
{ \partial T_{\eps}\over  \partial x} 
- {T_{\eps}} = f(x) \ ,
\eqno(17)
$$

\ni
where $\kappa(\xe)$ is a quasiperiodic function of $\xe$ with 
$\alpha_1 > \kappa(\xe) > \alpha_o > 0$ . $T_{\eps}$ tends
weakly in $H_1$ to $T_h$, where $T_h$ solves the homogenized
equation, i.e. $T_h$ is the solution of Eq. (17)  with
$\kappa(\xe)$ replaced by its harmonic mean:
$\kappa_h^{-1} \equiv <1/\kappa(\xe)>$. Convergence in $H_1$
means that 
${ \partial T_{\eps}\over  \partial x} $ converges 
weakly to ${ \partial T_{h}\over  \partial x}$  and 
this implies $T_{\eps}$ converges pointwise to $T_h$.

More generally, we consider a family of conductivities, $\kappa_{\eps}(x)$.
A general theory of the convergence of subsequences of the solutions to
the elliptic problem, Eq. (17), as $\eps$ tends to zero, has been developed,
and is termed `H' convergence. Since we consider sequences of oscillating
solutions, $T_{\eps}(x)$, weak convergence of the gradient of
$T_{\eps}(x)$ is to be expected. From linear operator theory, weak
convergence of all sequences $T_{\eps}(x)$,
where $L_{\eps}T_{\eps}(x)=f(x)$ and $f_{}(x)$ is arbitrary, 
in the Hilbert space $H_1$
is equivalent to the convergence of the operators, $L_{\eps}^{-1}$,
in the weak * operator topology. For one dimensional elliptic problems,
the main result of `H' convergence implies that if
$\kappa_{\eps}(x)^{-1}$ converges weakly to its average, then $T_{\eps}(x)$
converges weakly in $H_1$ to the solution of Eq. (17) with the harmonic
mean of $\kappa_{\eps}(x)$, i.e. 
$\partial_x \kappa_{h}\partial_x T_{h}(x)= f(x)$.

For our problem, we are interested in the weak convergence of 
solutions of Eq. (12) as $\Dh_{\eps}(x,q)$ is rescaled.
We wish to derive a similar effective equation for Eq. (12).
To do this, we must specify a family of possible rescalings of
$\Dh_{\eps}(x,q)$. The interesting rescalings of $\Dh_{\eps}(x,q)$  
are those rescalings where more and more resonant islands appear
and the domain has no remaining closed flux surfaces. In the
remainder of this article, we show that the physical rescalings
of the mode amplitudes, $\{ A_{mn} \}$, result in equations where
the solutions of $\Dh_{\eps}(x,q)$  converge weakly to an effective
equation with zero diffusivity.

We would like to homogenize $\Th_{\eps}(x,q)$ uniformly in $q$,
and then invert the Laplace transform to find an effective time
evolution equation for Eq. (6). However, Eq. (12) differs from the standard
case of Eq. (17), since $\Dh(x,q)$ has zeros and poles and is formally
small. Direct homogenization of Eq. (12) should still be possible when
$\Dh(x,q)^{-1}$ is integrable.

For extremely weak perturbations, the quasilinear diffusion term
is small away from the Doppler shifted resonances, i.e.
$\Dh \sim O(\eps_v^2 N_m )$. At each 
Doppler shifted resonance, $\xmn(q)$, a small resonance layer
of width $\delta x $ forms in which $\partial_x \Dh \partial_x$
becomes order one. The multiple scale analysis of Eq. (12)
separates into two cases: the ``isolated resonance"  regime
and the ``overlapping resonance" regime.

\bigskip
{\bf V. MULTIPLE SCALE ANALYSIS OF OVERLAPPING RESONANCES CASE}

The overlapping resonance case is easier to analyze and more physically
relevant. Thus we defer the isolated resonance case to Appendix E.
In the ``isolated resonance"  regime,
away from the Doppler shifted resonances, $\partial_x \Dh \partial_x \Th(x,q)$
is a small correction to the solution, $\Th(x,q) \sim T(x,t=0)/q$.
Near $\xmn(q)$, this ordering breaks down due to the resonance 
denominator in $\Dh$. 
As shown in Appendix D, the resonance layer width scales as
$\delta x \sim (\eps_v/\eps_L)^{2/3}/N_n$ for $q \sim O(1)$, and as
$\delta x \sim N_m^{1/3}(\eps_v/\eps_L)^{2/3}/N_n$ for $q \sim O(1/N_m)$, and
as $\delta x \sim (\eps_v/\eps_L)^{1/2}/N_n$ for $q =0$.


The treatment of the poles as isolated singularities requires
that the average distance between resonances exceed this resonance
layer width. Since the 
average distance between resonances is $1/N_mN_n$, the resonances decouple
if $ N_m^{3/2} \eps_v/\eps_L << 1$ for $q \sim O(1)$. 
For $q \sim O(1/N_m)$, the isolated resonance
criteria is $ N_m^2 \eps_v/\eps_L << 1$ and at $q=0$, the criteria  
is again $ N_m^2 \eps_v/\eps_L << 1$.

We now restrict our consideration to the overlapping resonance case, where
$1/N_nN_m$ exceeds the isolated resonance layer width.
There are approximately
$N_R \equiv N_mN_n \delta x$ resonances in the layer width.
In this case, {\it the
scalelength of rapid variation in $\Dh(x,q)$ and
$\Th$ is $\eps_m \equiv 1/N_nN_m$.
The appropriate maximal ordering is $\eps_m \sim \eps_v^2N_m$.}


We begin our multiple scale expansion by defining 
$y= \eps_m x$ and expanding $\Th_{\eps}$ as
$T_o + \eps_m T_1 +\eps_m^2 T_2$. 
Spatial differentiation has the expansion,
$1/\eps_m \partial_y + \partial_x$.
We expand the differential operator as
${\bf L}={\bf L_o}+ \eps_m {\bf L_1} + \eps_m^2 {\bf L_2} $ where
${\bf L_o}T \equiv \partial_y \Dh \partial_y T $ and
${\bf L_1}T \equiv \partial_x \Dh \partial_y T+
\partial_y \Dh \partial_x T -T$ and
${\bf L_2} \equiv \partial_x \Dh \partial_x$.

The zeroth order equation shows that $T_o$ is only a function
of $x$. The first order equation is
${\bf L_o}T_1+ {\bf L_1} T_o(x) = 0$.
The averaged first order equation yields $T_o = T(x,t=0)/q$. 

From the rapidly varying part, we have 
$T_1 = T_1(x) + \chi \partial_x T_o$, where 
$\chi$ is the solution of ${\bf L_o} \chi = -\partial_y \Dh$.
This equation may be integrated to yield
$ \left[  \partial_y \chi +1 \right] \equiv const/\Dh$. For $\chi$ to 
vary only on the fast spatial scale,
the constant must be the harmonic mean, $\Dbh$, of $\Dh$,  
i.e. $\Dbh^{-1}(q) \equiv < 1/ \Dh(x,q)>$. 
Thus $\partial_y \chi  = \Dbh^{-1}/ \Dh \ - \ 1$, and 
$\chi  = \int^y (\Dbh^{-1} /\Dh - 1)$. 

To have $T_{\eps_v,\eps_L}(x,q)$ converge to a limiting temperature,
$T_{lim}(x,q)$, pointwise, $\chi_{\eps_v,\eps_L}(x,q)$ must tend to
zero. Since $\partial_y \chi_{\eps_v,\eps_L}(y,q)$ is $O(1)$,
$\chi_{\eps_v,\eps_L}(x,q)$  will tend to zero pointwise if
and only if

$$ \int_a^x dy \left( \Dbh^{-1}(q)  /\Dh(y,q) - 1\right) 
\rightarrow 0 \ ,
\eqno(18) $$ 
i.e. weak convergence of $\Dh(x,q)^{-1}$ to $\Dbh(q)^{-1}$. 
Weaker hypotheses will result in weak convergence in the sense of
distributional limits. The separation of scales should be sufficient 
to ensure the existence of a weak limit. 
Since we are primarily interested in 
the limiting $T(x,t)$, a distributional
limit with respect to the Laplace parameter is probably sufficient
for our needs.

To the first order,  the temperature gradient 
vanishes at all resonance surfaces:
$$\partial_x  \Th(x,q) =  {\Dbh^{-1}\partial_x T(x,t=0)\over \Dh(x,q)\  q}
.\eqno(19OB) $$ 
The average of the second order equation is
$<T_1(x)> = \partial_x  < \Dh  \left[  1+ \partial_y \chi \right] >
\partial_x T_o$, which reduces to
$<T_1(x,q)> = \partial_x   \Dbh(q)  \partial_x T_o$.

The harmonic mean has logarithmic discontinuity across the real
$q$ axis whenever $\Dh(x,q)$ vanishes.     
When $\Dh(x,q)$ vanishes, our ordering assumption for the 
overlapping resonance regime is not strictly correct. In this case,
the correct ordering is locally identical with the isolated resonance regime.
Since the zeros of $\Dh$ generate {\it integrable} singularities,
the preceding asymptotic expressions should be valid.
Our multiple scale expansion correctly captures
the behavior of $\Th(x,q)$ at the zeros and poles of $\Dh(x,q)$.

At the double zeros, $\Dh$ vanishes too strongly to ignore the 
$- \chi$ term. In fact, the expression is meaningless since
$\Dbh(q)$ is undefined.
The presence of the second term in Eq. (12) removes the
nonintegrable singularities of $\Dh$.
At the double zeros, when $\partial_x^2 \Dh >>1$, 
the initial value basis solutions, $T_L$ and $T_R$, have
singularities of the form $1/(x-z)$. From the Green's function
representation, we see that $\Th$ has logarithmic singularities
away from the double zero.


When the two zeros are only
slightly displaced, we expect that the solution will approximate 
$x^{\lambda}$ by superimposing two logarithmic singularities.
The characteristic width for this strong interaction is  
$\zmn^{+}(q) - \zmn^{-}(q) = \delta x$. 
Surprisingly, this never occurs, since the resonance layer decreases
proportionally to $\partial_x \Dh \sim \zmn^{+}(q) - \zmn^{-}(q)$.
Thus this stronger $x^{\lambda}$ singularity manifests itself only
at the discrete frequencies where pairs of zeros coincide.
 

The preceding multiple scale expansion was based on the ordering,
$\Dh/\eps_m^{2} >> 1$. For large values of
the Laplace parameter, $q$, and spatially localized modes,
$A_{m,n}(x/n)$, 
the preceding expansion must be modified and the correct expansion
resembles that of Appendix E.
This may indicate extremely complex behavior for short times. 
 

\bigskip

{\bf VI. SINGULARITIES AND LONG TIME ASYMPTOTICS OF
THE OVERLAPPING RESONANCE REGIME}

To determine the long time behavior of $T(x,t)$, we 
analyze the singularities of $\Th(x,q)$.
We consider the homogeneous version of Eq. (12)
and let $\Th_L(x,q)$
be a solution which satisfies the left boundary condition
and let $\Th_R(x,q)$ satisfy the right boundary condition.
From the Green's function representation of $\Th(x,q)$, Eq. (15),
$\Th(x,q)$ has a Mittag-Leffler type expansion of the form:

$$\Th(x,q) = {T_f(x)\over q} +
\sum_{m,n,\pm} T_{pv}(z_{mn}^{\pm}(q),q)ln(x-z_{mn}^{\pm}(q))  +
\Th_S(x,q) \ ,
\eqno(20)
$$
where the subscript, $_{pv}$, denotes the principal value
and $\Th_S(x,q)$ 
has at worst  $xln(x)$ singularities. (See Appendix D.) 
The multiple scale expansion of Sec. V are of this form. 
At the isolated points where 
double zeros of $\Dh(x,q)$ occur, the logarithmic 
singularities in the second term should be replaced by
$(x-z_{mn}^{\pm}(q))^{\lambda}$ where $\lambda$ is calculated
in Appendix D.
For simplicity, we begin our analysis of the time evolution
of $T(x,t)$ by temporarily omitting the terms
arising from the degenerate resonances at
the double zeros of $\Dh(x,q)$.

Since we are interested in the slow evolution, we remove the rapid time
oscillations by filtering.
We define $T_W(x,t) \equiv T(x,t-s)*W(s)$ where $W(\cdot)$ is a
filter with bandwidth $c$.
The standard asymptotic expansion for long times is

$$ T_W(x,t) 
\equiv {1 \over \twp i}
\oint e^{iq\tb} \Th(x,q)\Wh(q) dq
=
T_f(x) + {-1 \over \twp i}
\oint e^{iq\tb} \Th(x,q)\Wh(q) 
dq \ .
\eqno(21) $$ 
As shown in Appendix C, {\it$\Th(x,q)$ is analytic for 
$q^2< min_{mn} A_{mn}/C_{mn} \sim 1/N_m$,}
except for the isolated pole at $q=0$. Thus
the second contour integral does not encircle these values
of $q$. 

If the bandwidth of the filter is $O(1/N_m)$ or smaller, 
the contour integral, representing the time
dependent part of  the solution, is exponentially small
and the time averaged state is effectively in the new saturated
state.

To examine the time dependent behavior, we need to use filter bandwidths
greater than $O(1/N_m)$.
To determine the leading order time dependent behavior, we integrate 
Eq. (21) by parts:

$$ T_W(x,t) =
T_f(x) + {-1 \over \twp i}
\oint {e^{iqt} \over i\tb} { \partial (\Th \Wh) \over  \partial q} 
(x,q) dq \ .
\eqno(22) $$

The $q$ scalelength for the variation of $\Th(x,q)$ is $1/N_m$.
If the bandwidth of the filter is larger, then 
$\Wh(q)$ may be removed from the $q$ derivative. Thus the leading order
long time behavior of $T_W(x,\tb)$ is

$$ T_W(x,\tb) \sim
T_f(x) + {1/\tb}\sum_{m,n,\pm} T_{pv}(x,q_{mn}^{\pm}(x))\Wh(q_{mn}^{\pm}(x)) 
exp(i\tb q_{mn}^{\pm}(x)) + O(1/\tb) \ .
\eqno(23)
$$
In the overlapping resonance case, the leading order term is 

$$T_{pv}(x,q_{mn}^{\pm}(x)) \sim <\Dh^{-1}(x,q_{mn}^{\pm}(x))>^{-1}
 \Dh_{pv}^{-1}(x,q_{mn}^{\pm}(x)) {T'(x,t=0)\over q_{mn}^{\pm}(x)}  \ .
\eqno(24)$$

In general, the remainder terms in the integrand satisfy 
$\partial_q \Th(x,q) \sim O(N_m \Th(x,q))$. 
Thus the size of the time dependent terms relative
to the $T_f(x) - T(x,t=0)$ scales as $(N_m/t)$.
This corresponds to temperature
oscillations persisting until a time of order $O(N_m)$ before phase 
mixing. {\it This time scale is also the time in which the argument
of the exponential, $exp(i\mu_{mn} t)$, becomes oscillatory on
the length scale of $1/N_mN_n$, the distance between harmonics.}
We cannot expect phase mixing to a Cantor set-like gradient
to dominate the time evolution on time scales which cannot distinguish
between these neighboring resonances.

The strength of the oscillations for times much longer than
the ideal time and approaching $t\sim N_m$ depends on the
behavior of $\Th(x,q)$ for small $q$. It is possible that the 
amplitude of the oscillations   decays considerably
in the time range $1<< t< N_m$.
Note that our estimate of the time to reach the new steady state is much shorter
than the naive diffusion time of $N_n/\eps_v^2$.
On the longer diffusive timescale, $\vb \tb/L > N_m$,
these rapidly decaying terms are only of slight interest.   
  
Our estimate of the size and timescale 
of the decay of the time dependent perturbation 
is unoptimized, since
we have not used
the more detailed asymptotic expansions of Sec. V and Appendix E
in our bound on the integrand $|\partial_q \Th|$.
In a future publication, we hope to apply our asymptotic 
representations of $\Th(x,q)$ to more precisely analyze the 
time evolution of $T(x,t)$ to $T_f(x)$.


The spatially averaged or slowly varying temperature, $T_S$,
satisfies

$$\partial_x <\Dh(q)> \partial_x \Th_S(x,q) - \Th_S(x,q) =
- T(x,t=0)/q \ ,
\eqno(25) $$
which corresponds to the evolution equation:

$$\partial_t T_S(x,t) = \partial_x \int_{s=0}^t K_h(t-s) \partial_x  
T_S(x,s) ds \ .
\eqno(26) $$

The averaged equation also approaches a steady state as $t \rightarrow \infty$.
Unfortunately, we are unable to simplify this spatially averaged
convolution equation further due to the complicated form of
$K_h(t)$. Eqs. (25)-(26) neglect the effect of the double zeros
of $\Dh(x,q)$, and they may modify the form of Eqs. (25)-(26).

We now present our understanding of the effect of
the degenerate resonances at the double zeros of $\Dh$
on the time evolution of $T(x,t)$.
In the case of overlapping resonances ($|\partial_x^2 \Dh(x,q)| >>1$),
$\Th(x,q)$ has a singularity of order $1/(x-z_{mn})$. So we
expect the oscillations at the double zero locations to damp
extremely slowly, if they damp at  all. In the isolated resonance
case ($|\partial_x^2 \Dh(x,q)| <<1$), the solutions  of
the homogeneous equation have $x^{-1/2+ic}$, where $|c|>>1$.
Thus we expect the oscillations of $T(x,t)$ at these points to phase mix
to zero at least as fast as $t^{-1/2}$.
However, the precise effect of the large complex exponent in the singularity 
requires further study.  
The conclusion that the time averaged state, $T_W$, reaches a new
Cantor set-like equilibrium is independent of the behavior at
the double zeros, since we can filter out this behavior.

In our simple equation, we are able to average spatially before
analyzing the time dependent behavior. In general, this type
of averaged equation will only occur when the spatial variation
is much stronger than the temporal variation.
  
\nsect
{\bf VII. DISCUSSION}

In conclusion, we have considered
the evolution of a passive scalar field
for a slowly varying
stratified medium which is convected in a turbulent sheared flow
within the quasilinear or weak coupling approximation.
On a nearly ideal time scale, $\tb \vb /L \sim N_m$,
 the fluid reaches a new saturated state.
The fluid remains in this modified equilibrium and does not  
evolve diffusively.
Our multiple scale expansion shows the final saturated state is

$$ T(x,t=\infty) \sim T(x,t=0) +
\left[ \int_a^x  ({\Dh_h(q=0)\over \Dh(x,0)}- \ 1.0 ) dx \right] 
\partial_x T_o +
\partial_x \Dh_h(q=0) 
\partial_x T_o \ .
\eqno(27)
$$
Near the resonances, $\delta T$ is order $\delta x_{q=o}$, and
$\partial_x T_o$ vanishes at each rational surface.
Since the growth of the harmonic, $\Tt$, is proportional to
$\partial_x T_o$, $\Tt$ saturates  
at each rational surface instead of growing linearly in time.

The resulting saturated state has a ``Cantor set" structure in the
gradient of the temperature. Such quasilinear Cantor set gradient
saturation has been found previously in self-consistent pressure 
gradient  driven turbulence in tokamaks$^{9-12}$. However 
the quasilinear ballooning mode calculations had an infinite number
of poloidal harmonics, but only one$^{9,10}$ or a small number of 
toroidal harmonics$^{11,12}$. In the present analysis, we have considered
situations where the density of toroidal harmonics tends to infinity
as well.

The long time solution of Eq. (6) is determined by balancing the transfer
of energy between the saturated state, $T_f(x)$ and the oscillations,
$\Ttw(x,t) \equiv T(x,t)-T_f(x)$. The rapidly varying part of 
the kernel, $\Kt(x,t)$, continuously excites temperature oscillations
from the equilibrium gradient. Initially, this transfer is large
due to the net coupling at the resonance surfaces. However, the
gradients at the rational surfaces rapidly flatten to zero and thereby
reduce the excitation to a small managable (balancable) level.

The standard weak turbulence theory of Galeev and Sagdeev$^3$ consists
of four succesive formal steps. First, in the weak coupling or
random phase approximation, the nonlinear interaction is truncated.
Second, the Markovian approximation replaces time history integrals
with the local time. Third, the continuum approximation replaces
the discrete sum over mode numbers, $\sum_{\kv}$ with an integral
over $k$ space, $\int dk$. Fourth, the rapidly varying diffusion
coefficient, $\kappa(x,t)$, is replaced by its weak limit, i.e. its
arithmetic average.
In this article, {\it we have shown that  
our model problem violates the last three formal steps, given the
truncated equations of the first formal step.}

In most turbulence formalisms, simplifications of the dynamics
are justified by hypothesizing statistical properties of the
the systems. If the hypothesized system properties are not good
approximations, the resulting analysis will fail.
Furthermore, these theories do not address the domain of validity
of these properties (such as the Markovian and continuum approximations).
In contrast, {\it our multiple scale expansion yields self-consistent 
estimates of the validity of our ``Cantor set" gradient saturated
state.}

Two other well known calculations, the Rechester-Rosenbluth theory$^{4,5}$
of particle diffusion and the estimate of the Kolmogorov exponentiation
length$^{15}$, $L_K$, both apply similar analysis techniques as Ref. 2. 
Thus the results of these calculations could easily be inconsistent
with a multiple scale expansion. In particular, nearly collisionless
particles follow the field lines almost perfectly, and therefore
will diffuse extremely slowly, in contrast to the collisionless
Rechester-Rosenbluth diffusion. 
The Kolmogorov exponentiation length calculation may still be
valid, since $L_K$ is a local property and the Markovian approximation
failure is related to the toroidal periodicity.

Our ordering implies that the fully nonlinear term is small, and
therefore may be neglected on the ideal timescale. 
For long times of order $N_m$, the
nonlinear term may significantly modify the asymptotic behavior of the
system. First, the fully nonlinear term will induce an intrinsically
nonlinear scattering and diffusion. Secondly and probably more
importantly, the nonlinear term will decorrelate the quasilinear
trajectories as they repeatedly encircle the torus. 

Much of the nonstandard behavior of our solution arises from the
infinite memory and oscillatory kernel of Eq. (6).
For long wavelength modes, the kernel is initially 
of order $O(\eps_v^2 N_mN_n)$.
However, on the velocity shear timescale, the nonresonant contribution
will decay away, and only a resonant part of the kernel, of order
$O(\eps_v^2 N_m)$, remains time asymptotically.
We note that the Markovian approximation corresponds
to replacing the oscillatory cosine kernel,
$ cos ({ \vbLd \umn(x){(s-\tb)\over \delta}} )$, by an exponentially
decaying kernel such as 
$ exp({ \vbLd \umn(x){(s-\tb)\over \delta}} ) $.
Thus the Markovian approximation cuts off the kernel after
the first oscillation and ignores the further oscillations.
Therefore it ignores the resonant contribution to the kernel.

The nonlinear
term should decorrelate this infinite memory and replace it
with an exponentially decaying kernel. If we make the common
assumption that the nonlinear decorrelation may be modeled
by a simple decorrelation time, $\tau_K$, the model equation 
becomes:
  
$$
{ \partial {T_o}\over  \partial \tb}=
\ckpd {\partial  \over \partial \xb}  
\sum_{m,n }^{}
A_{m,n}(x) \int_{s=0}^{\tb}
 cos ({ \vbLd \umn(x){(s-\tb)\over \delta}} ) 
exp( {(s-\tb) \over \tau_K(\eps_v,\eps_L)})
{\partial  T_o \over \partial \xb}(x,s)  ds   \ .
\eqno(28)
$$
The model of Eq. (28) is $not$ equivalent to assuming the random velocity
field has a slowly decaying autocorrelation function,
$<\vt(x,t),\vt(x,s)> \sim A_{mn}(x)exp((t-s)/\tau_k)$.
Equation (28) corresponds to the ensemble average
of the equations for various realizations of the probabilistic
velocity field with the given
autocorrelation function. However, the solution to the 
ensemble averaged equation is not necessarily the ensemble
averaged solution. Furthermore, we are interested in the ordering
where the nonlinear decorrelation time is long with respect to
the toroidal circulation time and Eq. (28) is probably only
physically reasonable in the opposite limit where $\tau_K<< L/\vb$. 

For large values of $\tau_K$,
the model nonlinear decorrelation of Eq. (28) allows the oscillations
to slowly decay in time. 
Although it is possible, we strongly doubtful that this
modified equation will have a $\tau_K$ independent limit which
diffuses according to the standard
weak turbulence theory.

{ In Ref. 15, a positive Kolmogorov length is calculated and this normally
implies chaotic behavior. Since we also assume a separation of
scales, diffusive behavior is to be expected. 
Clearly, we expect that the actual physical system will
experience diffusive evolution of some general type. However, the second
order truncation eliminates the nonlinear decorrelation of the
trajectories. The infinite memory of the kernel allows the trajectory to
circulate around the torus many times and average over the phase of
perturbing field. Thus
the effect of the modes phase mixes away except at the nearly resonant
surfaces and no diffusion results. 
Higher order calculations
are necessary to determine the strength of the nonlinear
scattering and the resulting transport.
  
In conclusion, the passive scalar problem differs from self-consistent
turbulence in two significant ways. First, in the Eulerian frame,
the autocorrelation time of the forcing is infinite. {\it Time
dependent random perturbations will tend to mix the fluid randomly
and eliminate infinite memory effects.} Thus the Markovian 
approximation is much more likely to be true with time dependent 
random perturbations or self-consistent fluctuations.

Second, the presence of a dissipation length would prevent
the quasilinear flattening of the gradients at the smallest 
wavelengths. If dissipation were to be included,
we still expect that the quasilinear flattening
would remain an important effect up to the dissipation scalelength.

Experimental observations indicate that fast particles have better
and not worse confinement than thermal particles$^{16}$. On the basis of
the Rechester-Rosenbluth calculations, the fusion community has inferred
that magnetic islands are not a significant source of transport in tokamaks.
Our results show that the collisionless limit of Rechester and Rosenbluth
is incorrect. The actual dependence of particle losses as a function
of parallel velocity is unknown. Therefore no inference on the presence
or absence of islands is possible.

A similar argument has been advanced in Ref. 16. In this calculation,
the net displacement of fast particles is gyrophase averaged to zero
due to the displacement of the particle trajectories from the magnetic
field line. The actual calculation in Ref. 16 is based on the extremely
questionable weak turbulence formalism of Ref. 2. We believe that a
multiple scale analysis of the particle trajectories also will show reduced
displacement due to gyrophase averaging. The strength of this effect is
presently unknown. 

Both our multiple scale analysis and the formal arguments of Ref. 16 find
greatly reduced diffusion due to phase averaging. Our mechanism requires 
the trajectory to make many circuits about the torus before the orbit
is decorrelated. The relative strength of toroidal phase averaging to
gyrophase averaging will depend on the mode spectrum and the decorrelation
mechanisms and deserves future study.

We now address the numerical estimates of diffusion
in overlapping island structures. Most of the calculations
report to be in either crude or good agreement with the
existing weak turbulence theory. We note that both our
analysis and the diffusive theory of Rosenbluth et al.
have an infinite number of free parameters, i.e. the  
spectrum $\vt$. Often the numerical simulations fix
the relative amplitudes of the modes and adjust only
the total amplitude, thereby testing only a small part
of the total ``spectrum" of problems (pun intended).

Furthermore, the quasilinear calculations are only valid
in the limit of many small overlapping modes. Naturally,
this limit requires much greater numerical resolution 
than a small number of large perturbations. Thus it
is quite possible that the majority of numerical simulations
have examined cases where quasilinear theory should not be valid.

The initial decay of the autocorrelation function is on the velocity
shear timescale; and this decay time is the basis for the quasilinear
diffusion coefficient. Inaccurate numerical schemes will observe this
initial decay, but may fail to resolve the saturation of the autocorrelation
function at times of order $N_m$. At this numerical resolution,
losses should correspond to the incorrect weak turbulence expression.

We note that the recent simulation by Duchs and Montvai$^{17}$
shows that the mean squared displacement saturates or 
evolves much more slowly than predicted by the standard
quasilinear diffusion coefficient.
This particular calculation is in strong qualitative agreement
with our multiple scale analysis.

\bigskip

\noindent
{\bf ACKNOWLEDGMENTS}

The final manuscript has benefitted from a critical reading
by D. Pfirsch.
Useful conversations with M. Avellaneda, 
R.V. Kohn, G. Papanicolaou, 
E.J. Riedel, H. Weitzner, D. Biskamp, S. Childress,
D. Duchs, E. Hameiri, M. Hugon, K. Imre, J. Krommes,  
A. Montvai, M. Ottaviani, 
R. White, L. Zakharov, and G. Zaslavskii
are gratefully acknowledged. 
This work was performed under U.S. Department of Energy, Grant No.
DE-FG02-86ER53223.

\newpage
{\bf APPENDIX A: PHYSICAL BASIS OF SPECTRUM RESCALING}

We assume the poloidal and toroidal mode numbers, $m$ and $n$, are large
and inversely proportional to a second small parameter, $\epsL$.
In our ordering, the $N_mN_n/2$ modes are of roughly equal magnitude,
and the amplitude scaling must satisfy 
$1/N_m^2N_n < \eps_v < 1/N_n\sqrt{N_m}$ for our quasilinear
overlapping resonance analysis to apply. Our results should be
easily extendable to a much larger class of mode spectra. The
lower bound, $1/N_m^2N_n < \eps_v$
generalizes to the overlap criteria
$\sum_{m,n}\left({\eps_v^2 |\vt_{mn}|^2 \over \umn'(x_{mn})^2}
\right)^{1/4} > 1$.
The upper bound, $\eps_v < 1/N_n\sqrt{N_m}$,
generalizes to
$\sum_{m,n,|x-x_{mn}|<1/N_n}{\eps_v^2 n^2|\vt_{mn}|^2 } << 1$.
This upper bound is crude since it is a simple estimate of the relative 
strength of the fully nonlinear term relative to the quasilinear term.

Our analysis does require that the mode amplitudes have explicit cutoffs
at $N_m$ and $N_n$. Furthermore, our results on the timescale for the
decay of the time dependent oscillations is in terms of the mode cutoff.
For more general expressions about the temporal evolution, the multiple
scale expansions need to be used directly in Eqs. (21)-(23).

Mode spectra arise from two classes of phenomena: externally driven 
perturbations and slowly evolving turbulent fluctuations 
with saturated magnetic islands. We now describe the characteristics
of each class of spectra. Although the spectra have different radial
scalelengths, their effective properties are nearly identical due
to the localizing effect of resonant denominators.

Reference 2 considered the case 
where the perturbing field is externally
driven and therefore $A_{mn}$ were treated as constants. 
Due to the presence of resonant denominators of the form,
$1/(n\vzx -m\mux-\omm)$, the perturbations are effectively localized
radially about the Doppler shifted resonance. The localization
width is proportional to $\epsL$.

In turbulent situations characterized by a nonlinear cascade of 
energy, we expect all harmonics within a given wavelength range
with rational surfaces to be excited. The higher mode numbers usually
have decreasing amplitudes. This corresponds to a sector in
the $(m,n)$ plane with the number of excited modes proportional
to $1/\epsL^2$. Thus the number of excited poloidal mode
numbers, $N_m$, and excited toroidal mode numbers, $N_n$, both
scale as $1/\epsL$. There are two cases of interest: short wavelength
and long wavelength perturbations. 
Short wavelength modes decay radially on the
scalelength of $\eps_L$, and therefore have an effective mode number
density of $N_m$, and an 
energy density of excited modes
is proportional to $\eps_v^2 N_m$.   

Long wavelength modes decay on the macroscopic
scalelength  and therefore have an effective mode number 
density of $N_mN_n$, and an 
energy density of excited modes
is proportional to $\eps_v^2 N_mN_n$.
However, the underlying velocity shear causes
the effect of the long wavelength modes to phase mix away except
in a small resonance layer, where $\kv \cdot \vv_o(x)$ is small. 
The timescale for the nonresonant terms to decouple is the
velocity shear time, $1 / \partial_x \mu(x)$. The quasilinear
diffusion coefficient of Ref. 2 is essentially the Kubo diffusion
coefficient, using $<\vt,\Ttw>$ in place of $<\vt,\vt>$  
with this nonresonant decorrelation time. {\it 
However the long time behavior is dominated by the resonant layers
and this modified ``nonresonant" Kubo formula is not applicable.}
Since the long wavelength   
modes are resonant in only a small region of order $1/N_n$, 
the resonant energy density is smaller, $O(\eps_v^2 N_m)$.

However, 
we also wish to consider cases
where $N_n \sim 1/\epsL$ and $N_m$ is a fixed subsidiary large
parameter. In this case,
the energy density of the the excited modes
is independent of $\epsL$ and  proportional $\eps_v^2$.
Therefore we specifically distinguish $N_n$ and $N_m$
in order to treat both cases simultaneously. 

We assume that the velocity shear, $\mu'(x) \equiv {a\over L} v_{\theta}'$,
is order one. Our transport becomes small when 
$cos(\umn(x)t)$ is oscillatory on the scalelength of the mode
separation, $1/N_m N_n$. If $\mu'(x)$, the $x$ coordinate, and
time were rescaled so that
$\umn(x)\times \ t$ were small, a different limiting equation would result.  

A number of articles in nonlinear dynamics have considered the
$1 \ 1/2 $ degree of freedom Hamiltonian system analogous to the
passive scalar flow field: $\vv_o(x) = \zh + x\thh$ and
$\vvt ={1\over \eps_v} \vt \xh 
\sum_{m=-\infty}^{+\infty} cos(m\theta - cz +\phi_m)$. This flow field
is a combination of a single chain of resonant perturbations of large
(not small!) amplitude.
We rescale this system by changing variables:
$x_{new} \equiv \eps_v^2 x_{old}$ and thereby show that these Hamiltonian
system correspond to scaling the velocity shear to infinity as
$\partial_x \vv_o(x) \sim 1/ \eps_v^2$.

\newpage
{\bf APPENDIX B: LAGRANGIAN TRUNCATION}

We now perform an alternative expansion in  Lagrangian coordinates.
This expansion is motivated by the results of Kraichnain$^{16}$ that
Lagrangian direct interaction approximations are necessary to
recover the famous Kolmogorov inertial range.
We let $\xv(t,\xv_o)$ denote the position of a particle
which was located at $\xv_o$ initially. We continue to denote the
stratified direction by the scalars, $x$ and $v$ and assume that
the equilibrium velocity, $\vv_o$ and the wave vectors, 
$\kv \equiv (0,m/a,n/L)$,
lie in the stratified plane. We
expand $\xv(t,\xv_o)$ as $\xv^o(t,\xv_o)+ \eps_v\xv^1(t,\xv_o) 
+\eps_v^2\xv^2(t,\xv_o)$ .

The particle trajectory satisfies

$$\partial_t{\xv}(t,\xv_o) = \vv_o(x) + 
\eps_V\sum_{\kv} \vv_k(x) e^{i\kv\cdot \xv} \ .
\eqno (B1)
$$
The zeroth order solution is $\xv^o(t,\xv_o) = \xv_o + \vv(x_o) t$. 
The next order equation is
 $$\partial_t{\xv}^1(t,\xv_o) = 
{\partial \vv_o(x_o) \over \partial x} x^1(t,\xv_o)  +
\sum_{\kv} \vv_k(x_o) e^{i\kv\cdot\xv^0}
. \eqno (B2) $$
We integrate the $x$ component of Eq. B2, $\partial_t{x}^1(t,\xv_o) = 
\sum_{\kv} \vv_k(x_o) e^{i\kv\cdot\xv^0}$, and then substitute
$x^1(t,\xv_o)$ into the $y$ and $z$ components to yield:

${\xv}^1(t,\xv_o) = $
$$ \sum_{\kv} e^{i\kv\cd\xv_o}  \left[ 
\vv_k(x_o) 
\left( { { e^{i\kv\cd\vv_o(x)t} -1}\over i\kv\cd\vv_o(x) } \right)
+ {\partial \vv_o(x_o) \over \partial x} v_{x,k}(x_o) 
\left( { { e^{i\kv\cd\vv_o(x)t} -1 - i\kv\cd\vv_o(x) t }\over 
-(\kv\cd\vv_o(x))^2 } \right) 
\right] \eqno (B3) $$
The second order equation is
 $$\partial_t{\xv}^2(t,\xv_o) = 
{\partial \vv_o(x_o) \over \partial x} x^2(t,\xv_o)  +
{\partial^2 \vv_o(x_o) \over \partial x^2} {x^1(t,\xv_o)^2\over 2}  +
\sum_{\kv} e^{i\kv\cd\xv^0} \left( \partial_x \vv_k(x_o) x^1 +
\vv_k(x_o) i\kv\cdot\xv^1 \right)
\eqno (B4)
$$
The normal component reduces:

$\partial_t{x}^2(t,\xv_o) = $
$
\sum_{\kv,\kv'} e^{i(\kv+\kv')\cdot (\vv_o(x)t+\xv_0)} 
\left( v_{k',x}(x_o) \partial_x v_{k,x}(x_o) +
v_{k,x}(x_o) i\kv \cdot \vv_{k'}(x_o) \right)
\left( { {1 - e^{ -i\kv'\vv_o(x)t} }\over i\kv'\cd\vv_o(x) } \right) 
$
$$ +
\sum_{\kv,\kv'} e^{i(\kv+\kv')\cdot (\vv_o(x)t+\xv_0)} 
i {\partial \kv\cd\vv_o(x_o) \over \partial x} 
v_{k,x}(x_o) v_{k',x}(x_o) 
\left( { {1 - (1+i\kv'\cd\vv_o(x)t)e^{ -i\kv'\cd\vv_o(x)t} }
\over (\kv'\cd\vv_o(x) )^2 } \right) 
\eqno (B5)
$$

The time independent part corresponds to $\kv' \equiv -\kv$ and
is identically zero due to incompressibility.
For a finite number of modes, these expressions imply that the 
mean squared displacement, $<|x(t,x_o)-x_o|^2>$, is bounded
except at the rational surfaces. This implies that there
is no long time diffusion unless the number of resonance
surfaces tends to infinity.

If we make the random phase assumption, the crossterms phase average
away and we find the autocorrelation time is
 
 $$<|x(t,x_o)-x_o|^2>
= 2 \ \sum_{\kv} {|v_k(x_o)|^2  \over |\kv \cdot \vv_o(x_o)|^2 }
\left( { {1 \ - \ cos({\kv\cdot\vv_o(x)t}) } } \right).
\eqno (B6)
$$

Thus the Lagrangian truncation also predicts zero diffusion away
from the rational surfaces. Our saturated state has 
$\partial_x T(x_{mn}) =\ 0$ at all rational surfaces and therefore 
is compatible with the Lagrangian result.
In the opposite limit of small shear time, $\kv\cdot\vv_o(x)t<<1$,
we find wave-like transport:
$<|x(t,x_o)-x_o|^2>
= \left( \sum_{\kv} |v_k(x_o)|^2\right) t^2$.

\newpage
{\bf APPENDIX C: SMALL Q BEHAVIOR}
 
For small values of $\delta q$, corresponding to large times,
the two poles, $x_{mn}^{\pm}(q)$ coalesce near
$x_{mn}$. We now examine $\Dh(x,q)$ in a small  
neighborhood of $x_{mn}$. To separate the resonant and nonresonant
parts of $\Dh$, we define
$C_{mn} \equiv \Dh(x,q) - A_{mn}(x)/ (\umn(x)^2-q^2)$ .
The scalelength for variation of the resonant part, 
$A_{mn}(x)/ (\umn(x)^2-q^2)$, is 
$2m\mu'\umn(x)/ (\umn(x)^2-q^2)$. This is large near the poles
and order $N_n$ away from the resonance surfaces. 


In a small neighborhood of $x_{mn}$, only the
resonant denominator variation is important and
we treat the nonresonant contribution,
$C_{mn}(x,q)$ as a constant, $C_{mn}$. Thus we define
$C_{mn} \equiv \Dh(x,q) - A_{mn}(x)/ (\umn(x)^2-q^2)$    
and suppress the $x$ and $q$ dependencies in $C_{mn}$ and $A_{mn}$. 
Note that $\Dh(x,q)^{-1}$ is approximately 
$$ \left( C_{mn} + { A_{mn}  \over 
  {\umn^2(x)}-q^2 } \right)^{-1} 
=
1/C_{mn} - \ \ \ \ 
\eqno(C1)
$$
$$
 {A_{mn}\over 2 C_{mn}^2\sqrt{ q^2 - A_{mn} /C_{mn} }} 
\left( {1 \over {\umn(x)}-\sqrt{ q^2 - A_{mn}  /C_{mn}  }} -
{1 \over \umn(x) + \sqrt{ q^2-A_{mn}/C_{mn} }} \right) \ .
$$

Thus if $q^2>A_{mn}/C_{mn}$, $D(x,q)$ has two zeros, $z_{mn}^{\pm}(q)$,
nestled between the poles, $x_{mn}^{\pm}(q).$  We denote the inverse
function by $q_{mn}^{\pm}(x).$ 
Note that the jumps in the 
values of the logarithmic terms at the branch points, $z_{mn}^{\pm}(q)$, 
do not cancel. At the zeros, $z_{mn}^{\pm}(q)$, 
$$ \partial_x\Dh(z_{mn}^{\pm},q) =
\sum_{m'n'} 2{\mu_{m'n'}(z_{mn}^{\pm})\mu_{m'n'}'A_{m'n'}
\over(\mu_{m'n'}^2-q^2)^2} 
+ 
{A_{m'n'}'(z_{mn})\over (\mu_{m'n'}^2-q^2)} 
\eqno (C2) $$
$$\sim 2 \umn(z_{mn}^{\pm})\umn'C_{mn}^2/A_{mn} \ . $$
In our ordering $A_{mn}/C_{mn}$ is order $1/N_m$, which is a subsidiary
small parameter, thus $\partial_x\Dh(z_{mn}^{\pm},q)$  
is $O(m\eps_v^2 N_m^2)$. 
Since the jump in 
${ \partial \Yh\over  \partial x}$ is inversely proportional to
$\partial_x\Dh(z_{mn}^{\pm},q) $ and there are approximately 
$2N_n N_m$ zeros, {\it the total contribution to the change in $\Th(x,q)$
from the sum of all the resonances is
exactly the same order as the nonresonant
contribution.}
 
At 
$q^2= A_{mn}/C_{mn}$, the two zeros coalesce and $\Dh(x,q)$ is approximately
${C_{mn} \umn(x)^2 \over \umn(x)^2 -q^2}$. 
The solutions of the homogeneous equation, $T_L$ and $T_R$,
are highly oscillatory when $|\umn(x)|<q$.
$\Th(x,q)$ has singularities of order
$(x-\xmn)^{\lambda}  $ where $\lambda =(-1 \pm \sqrt{ 1-4/g})/2$ where
$g\equiv C_{mn}(m\mu')^2 / q^2$. 
In our ordering with clearly separated poles, $g$ is  
$O(\eps_v^2 N_m^2 /\eps_L^2 )$, which is much less than one.  

For 
$q^2<A_{mn}/C_{mn}$, Eq. (12) has no singularities except a simple
pole at $q=0$. The pole generates a steady solution, $T_f(x)$,
corresponding to $\Th(x,q) \sim T_f(x)/q$. 
Since $\Th_{\eps}$ is analytic in $q$ for $0<q^2<A_{mn}/C_{mn}$, 
this range of $q$ does not contribute to the time evolution.

For $|q|^2 >max_{x,m,n} \umn(x)^2$, $\Dh(x,q)$ is negative and
does not vanish.
Thus the continuous spectrum exists only in the range 
$ 1/ \sqrt{N_m} < |q| < N_m.$ 

\newpage
{\bf APPENDIX D. SINGULARITIES OF $\Th(x,q)$: REAL AND REMOVABLE}

The long time asymptotic behavior of $T(x,t)$ depends on the
regularity of $\Th(x,q)$ which in turn depends on the behavior of 
$\Dh(x,q)$. 
Due to the positivity of $A_{m,n}$, $\Dh(x,q)$ has zeros and poles only
on the real $q$ axis. 
$\Kh(x,q)$ can be decomposed into pairs of poles,

$$\Kh(x,q) \equiv \sum_{m,n} A_{m,n}(x) \left( 
{1 \over \mu_{m,n}(x) - q  }
- {1 \over \mu_{m,n}(x) + q  }\right)  \ .
\eqno (D1)
$$
\ni 
As $q$ approaches the real axis, we can split $\Kh(x,q)$  
into a real principal part, $\Kh_{PV}(x,q)$, plus a sum of
delta functions:
$lim_{ q_I \rightarrow +0} \Kh(x,q) =$
$$
 \Kh_{PV}(x,q)+
\pi i \sum_{m,n} \left( 
{A_{m,n}(x^+_{m,n}) \over\mu_{m,n}'(x)}  \delta (x -x^+_{m,n}(q))  
+ {A_{m,n}(x^-_{m,n})\over\mu_{m,n}'(x)}   
\delta (x -x^-_{m,n}(q))  
\right)  \ .
\eqno (D2)
$$
    
Furthermore, for small $q$, $\Kh_{PV}(x,q)$ tends to zero
(the real parts of the denominators cancel). The standard
derivation$^2$ of quasilinear diffusion averages the limiting
$\Kh(x,q)$ spatially. The flaw in this analysis is that {\it the 
limit of the solution is not the solution of the limit.}
In fact, the poles of $\Dh(x,q)$ generate removable singularities.
The zeros of $\Dh(x,q)$ generate actual singularities in 
$\Th$. 
The limiting potential, $V_{lim}(x,q_R)$, in Eq. (14) satisfies
$$ lim_{ q_I \rightarrow +0} V(x,q) =
 V_{PV}(x,q)+
\pi i \sum_{m,n} \left( 
{V(z^+_{m,n}) }  \delta (x -z^+_{m,n}(q))  
+ {V(z^-_{m,n})}
\delta (x -z^-_{m,n}(q))  
\right), 
\eqno (D3)
$$
where $V(x,q)\equiv \Dh(x,q)^{-1}$, and the poles of $V(x,q_R)$  
are the zeros of $\Dh(x,q_R)$.
When the potentials, $V_{\eps}(x,q)$, are uniformly
bounded from above and below, $0<c_l<V_{\eps}(x,q)<c_u$, 
the limiting equation for 
$\Yh(x,q) \equiv \Dh(x,q) \partial_x \Th(x,q)$ exists, and the
solution of the limit is the limit of the solutions.
In our case, the potentials, $V_{\eps}(x,q)$ are not 
bounded from above and below, but are integrable except at
the double zeros of $\Dh(x,q)$. Since the multiple scale 
expansion appears to only require integrability, we believe 
similar convergence theorems will hold.

As the Laplace parameter, $q$, is varied, the positive poles,
$x^+_{m,n}(q)$, move to $+\infty$ with speed,      
$\partial_q x^+_{m,n}(q) = 1/ \mu_{m,n}'(x)$.
On its path, each pole will collide with a large number of other
poles. 
As $q$ varies, the Doppler shifted poles are displaced and occasionally
coincide. 
For $q \ne 0$, $\Dh(x,q)$ remains with a $1/x$ singularity.
Thus only the strength and not the order of the singularity is increased.

Between each two adjacent poles, there will be one, two
or no zeros. If the adjacent poles have the same parity,
there will be an odd number of zeros, generically, one.
Between adjacent poles of opposite parity, 
there will be an even number of zeros. When the poles are
infinitesimally close together, there are no zeros.
However as the poles separate, a double zero may form at a
critical value of $q$. As the distance further increases,
the double zero will divide into two separate isolated zeros.
In Appendix C, we show that near $q=0$, the zeros of $\Dh(x,q)$
evolve from no zeros to a double zero to two single zeros
as $q$ increases.

The single and double {\it zeros generate 
spatially localized temperature oscillations which decay algebraically
in time.}
We now examine the {\it local solutions} of Eq. (12) at the
poles and then at the zeros. We note that $\Dh(x,q)$ has double
poles only at $q=0$. Thus the generic case is when
$\Dh(x,q)$ is proportional to $1 / (x - x_o(q) )$
at each resonance surface.
From Eq. (15), we see that
$\Yh(x,q) \equiv \Dh(x,q) { \partial \Th\over  \partial x} (x,q) $
is locally analytic in $q$ at the poles and double poles of $\Dh$. 
Thus $\partial_x \Th(x,q)$ must vanish at $x_{mn}(q)$. 
At the double poles at $q=0$, both
$\partial_x \Th(x,q)$ and
$\partial_x^2 \Th(x,q)$ vanish.  
Thus the poles of $\Dh(x,q)$ generate
removable singularities.

The zeros of $\Dh(x,q)$ generate singularities of the first kind in
$\Th(x,q)$.
We now examine the behavior of the solutions of the homogeneous equations 
near a zero, $z_o(q)$, of $\Dh(x,q)$. We let $c_0(q)^{-1} =
{ \partial \Dh \over  \partial x} (z_o(q),q)$.
Near the zero, there is a  solution,
$\Th_A(x,q)$, which is locally analytic in $q$
with $\Th_A(z_o(q),q)=1$ and ${\partial \Th_A \over \partial x}
(z_o(q),q)= c_o(q)$.
The corresponding locally analytic solution of Eq. (14) is
$\Yh_A(x,q)$ with $\Yh_A(x,q) \sim  x-z_o(q) + c_o(x){(x-z_o(q))^2 \over 2}$.

The second solution,
$\Th_S(x,q)$, has a logarithmic singularity of the form
$\Th_S(x,q)= ln(x -z_o(q))\Th_A(x,q) + \Th_C(x,q)$ where
${\Th_C }
(z_o(q),q)= 0$.
The corresponding local solution of Eq. (14) is
$\Yh_S(z_o(q),q)= ln(x -z_o(q))\Yh_A(x,q) + \Yh_C(x,q)$,
where $\Yh_C(z_o(q),q)= 1 / c_o(q)$ and $\Yh_C(x,q)$
is locally analytic.
We note that any solution satisfies the jump condition:
${ \partial \Yh\over  \partial x}
|_{z(q)-\eps}^{z(q)+\eps} = \pm \pi i c_o(q)  \Yh(z_o(q),q)$.

At a double zero of $\Dh$, there are solutions, 
$\Th(x)_{\pm} \sim z^{\lambda}$ where 
$\lambda = -1/2 \pm \sqrt{1/4 - 1/g} $ and
$g \equiv \Dh''(x)$. Thus if $0<g<<1$, then the
singularities are order $z^{-1/2}$. For 
$\Dh''(x) >> 1$, the singularity approaches $z^{-1}$. 
Overlapping resonance layers correspond to
the case $\Dh''(x) >> 1$.

A second aspect of the singularities of $\Th(x,q)$ is the
thickness of the resonance layers about the zeros and poles
of $\Dh(x,q)$. The resonance layer is the subdomain where
the behavior of $\Th(x,q)$ is dominated by the singularity.
We examine the intrinsic layer width for poles, double
poles, zeros and double zeros. 

For the case of a single isolated resonance,
the resonant layer width scales as $(n\delta x)^3 \sim
 (\eps_v/\eps_L)^2 (n/m\mu')A_{mn}/(\umn(x)+q)$.
In our evaluations of the asymptotic size of terms, we need
estimates of $O({1 \over \mu_{mn}+q})$ near a resonance of
$\mu_{mn}-q$. This naturally depends on our choice of $q$.
For small $q$, $q = O(1/N_m)$, the denominator is $O(N_m)$.
However, { the inverse Laplace transform involves 
integrals, $dq$, and the small $q$ are downweighted in the 
integration due to their small measure.} We find the $q$ weighted 
average scaling of the terms is not significantly altered
by the modified scaling at $q<<1$.

When $q$ is $O(1)$, the layer width scales as
$(n\delta x)^3 \sim (\eps_v/\eps_L)^2 $.
For $q \sim 1/N_m$, 
the layer width is
$(n\delta x)^3 \sim (\eps_v/\eps_L)^2 N_m$.
Finally, at $q=0$, all poles are double poles and
the resonance layer width scales as $(n\delta x)^4 \sim
 (\eps_v/\eps_L)^2 (n/m\mu')^2$.  
The total area occupied by resonance layers scales roughly
as $N_mN_n \delta x $. For $q \sim O(1)$,
$N_mN_n \delta x \sim (\eps_v/\eps_L)^{2/3} N_m^{}$.
For $q \sim O(1/N_m)$,
$N_mN_n \delta x \sim (\eps_v/\eps_L)^{2/3} N_m^{4/3}$.
For $q = 0$,
$N_mN_n \delta x \sim (\eps_v/\eps_L)^{1/2} N_m^{}$.

At the zeros, $\zmn(q)$,
of $\Dh$, the boundary layer thickness is much smaller,
$\delta x \sim
\partial_x\Dh(z_{mn}^{\pm},q) =2 \umn(z_{mn}^{\pm})\umn'C_{mn}^2/A_{mn}$,
which is $O(m\eps_v^2N_m^2  \sqrt{ q^2-A_{mn}/C_{mn} } )$. 
Note this is the boundary layer thickness of the homogeneous equation
and not the inhomogeneous equation.

For double zeros, the extent of the subdomain where resonant  
behavior dominates depends on the ``exterior" solution and
cannot be determined {\it a priori}.

\newpage
{\bf {APPENDIX E: MULTIPLE SCALE ANALYSIS OF THE ISOLATED }
{RESONANCES REGIME}}

In this appendix, we find multiple scale solutions to Eq. (12) in
the isolated resonance case. The expansions can the be used in Eqs. (21)-(23)
to derive the leading order time dependent asymptotics. In practice,
we use the isolated resonance expansion only to show that the solutions
do not become pathologically large or singular. With some additional
work, this expansion could be incorporated into Sec. VI.
To solve the zeroth order equation, we used the W.K.B. expansion
with turning points.

Due to the smallness of $\Dh(x,q)$, 
the standard multiple scale ordering$^{6-8}$ must 
be slightly modified. For a maximal ordering,
we assume $\eps_v^2 N_m \sim \eps_L^2$. 
This ordering enables us to correctly treat the resonances of $\Dh$.
We let $y= \eps_L x$ and expand $\Th_{\eps}$ as
$T_o + \eps T_1 +\eps^2 T_2$. Spatial differentiation has the expansion,
$1/\eps_L \partial_y + \partial_x$.
We expand the differential operator as$
{\bf L}={\bf L_o}+ \eps {\bf L_1} + \eps^2 {\bf L_2} $ where
${\bf L_o}T \equiv \partial_y \Dh \partial_y T -T$ and
${\bf L_1} \equiv \partial_x \Dh \partial_y +
\partial_y \Dh \partial_x$ and
${\bf L_2} \equiv \partial_x \Dh \partial_x$.
${\bf L_o}$ contains both  slowly and rapidly varying parts due
to the presence of the identity operator. The slowly varying
part of ${\bf L_o}$ satisfies $<{\bf L_o}T> = -<T(x,t=0)>$. 

The zeroth order equation reduces to $T_o = T(x,t=0)/q$. To
next order, we have $T_1 = \chi \partial_x T_o$, where 
$\chi$ is the rapidly varying solution of ${\bf L_o} \chi = -\partial_y \Dh$.
$\chi(x,q)$ will have logarithmic singularities at the zeros
of $\Dh(x,q)$ and $(x-z)^{\lambda}$ singularities at the double zeros.

Within this resonance layer, the full equation,
${\bf L_o} \chi = -\partial_y \Dh$, must be solved.
Since $\partial_x \Th(\xmn(q),q) =0$, 
$\partial_y \chi(\xmn(q),q) = -1$ and thus near $\xmn(q)$,
{\it $\chi$ is order $\delta x$, 
independent of the value of $\eps_v \sqrt{N_m} / \eps_L$!}  
Within the layer,
$\partial_q\chi(x,q) \sim \partial_y \chi(x,q) \partial_q \xmn(q)$,
and therefore $\partial_q\chi(x,q)$ is order $1/m\mu'$.  

Away from the poles, $\xmn(q)$, and  
assuming $\delta q \sim O(1)$, $\Dh$ is $O(\eps_v^2 N_m )$,
$\partial_x \Dh$ is $O(\eps_v^2 N_m / \eps_L)$, 
and $\partial_q \partial_x \Dh$ is $O(\eps_v^2 N_m / \eps_L)$. 
If $\partial_y ln(\chi)$ is O(1),
$\partial_y \Dh \partial_y \chi$ is order $(\eps_v/\eps_L)^2 N_m$
smaller than $\chi$ away from the the resonances, $\xmn(q)$.

We would like to expand $\chi(x,q)$ in powers of $(\eps_v/\eps_L)^2 N_m$
away from the poles.
Between the poles of $\Dh(x,q)$, a particular solution of the inhomogeneous
equation is
$\chi_p = \partial_y \Dh$, accurate to order $O( (\eps_v/\eps_L)^2 N_m)$. 
Away from the poles of $\Dh(x,q)$, the solution would appear
to converge to 

$$
T(x;q) = {T(x,t=0) 
+\partial_x \Dh(x,q) \partial_x T(x,t=0) +...\over q} \ .
\eqno (E1)
$$

This formal solution clearly fails near the poles of $\Dh(x,q)$
where the actual solution has at worst an $(x-\xmn(q))ln(x-\xmn(q))$
singularity, as shown by the local analysis and the Green's function
representation. 

In the Green's function of Eq. (15), $T_L(x,q)$ is
a solution which satisfies the left boundary condition
and $T_R(x,q)$ satisfies the right boundary condition.
Replacing $T_{L,R}$ by $\partial_{\xi} \Dh \partial_{\xi} T_{L,R}$
in Eq. (15) and integrating by parts yields


$$ A(q)T(x;q) = {A(q)T(x,t=0)\over q} - $$
$$
T_R(x;q) \int_{a} ^{x} ( \Dh(\xi,q) \partial_{\xi} T_L(\xi,q) ) 
\partial_{\xi} f(\xi,q) d\xi 
- T_L(x,q) \int_{x} ^{b} \Dh(\xi,q) (\partial_{\xi} T_R(\xi,q) ) 
\partial_{\xi} f(\xi,q) d\xi  \ ,
\eqno (E2)$$
where $A(q) \equiv \Dh(x;q) W(T_L,T_R,x)$ is independent of x
and $f(x,q)=-T(x,t=0)/q$. 

Away from the zeros and poles of $\Dh(x,q)$, $T_L(x,q)$
and $T_R(x,q)$ can be represented in terms of the W.K.B. expansion$^{19-21}$:
$$T_{\pm}(x,q) =c_{\pm}(q) exp(\pm{\Phi(x,q)\over\eps_L})
\sum_k \eps_L^k p_k^{\pm}(x,q)  \ ,
\eqno (E3)
$$  
where $\partial_x \Phi(x,q) \equiv 1/\sqrt{\Dh(x,q)}$ and
$p_o^{\pm}(x,q) = {\Dh(x,q)}^{-1/4}$.
The sign of $q_I$ determines the branch cut of $\Phi$ in the
W.K.B. representation of $T_L$ and $T_R$ and this
breaks the analyticity of $\chi$ with respect to $q$ .

In regions where
$\Dh > 0$, the solutions grow exponentially,  
and this exponentially growth localizes the kernel of
the Green's function to a neighborhood of $x$. 
When $\Dh < 0$, the solutions are oscillatory, and the
entire subinterval 
can contribute to
the kernel, as well as neighborhoods of the two adjacent
transition points. 

A further integration by parts yields

$$ A(q)T(x;q) = {A(q)T(x,t=0)\over q} + B(q) + $$
$$
T_R(x;q) \int_{a} ^{x} T_L(\xi;q) 
\partial_{\xi} \Dh(\xi,q)\partial_{\xi} f(\xi;q) d\xi 
+ T_L(x;q) \int_{x} ^{b} T_R(\xi;q) \partial_{\xi} \Dh(\xi,q) 
\partial_{\xi}f(\xi;q) d\xi  \ ,
\eqno (E4)$$
where $B(q) \equiv \Dh(x;q) (\partial_{x}f(x;q) )
T_L(\xi,q)T_R(\xi,q)|_{\xi=a}^b$
is a boundary term, independent of $x$ and
$exponentially$ $small$. 
This procedure of replacing $T_{L,R}$ by 
$\partial_{\xi} \Dh \partial_{\xi} T_{L,R}$
in Eq. (15) and integrating by parts may be repeated successively to
formally rederive the expansion in Eq. (E1). Again the derivation is
flawed due to the poles of $\Dh$.  Away from the transition 
points, in regions where $\Dh > 0$, the effect of the poles of $\Dh$ is
exponentially small and may be neglected. 

Thus we need to consider only regions where $\Dh$ is negative and
regions near a zero or pole of $\Dh$.
Equation (16) shows that the time evolution
of $T(x,t)$ is completely determined by $Imag(\Th(x,q_R^-)$. Since
$\Th_L$ and $\Th_R$ are analytic at the
poles of $\Dh$, the
poles of $\Dh$ do not generate imaginary $\Th(x,q)$.
Therefore points, $(x,q)$, which lie on anti-Stokes lines between two
adjacent poles are well represented by Eq. (E1), and
do not contribute to the time evolution according to
Eq. (22).
If the point, $(x,q)$, lies on anti-Stokes lines between two
zeros of $\Dh$, the expansion of Eqs. (E1) and (E4) converges.
Thus we need only evaluate the Green's function when 
the point of interest, $x$, lies in an interval of negative
$\Dh(x,q)$ which is bordered by one pole and one zero.

Using the method of stationary phases, it can be shown that
the only contributions to the integral representation
of Eq. (E4) occur at the adjacent transition points and at the point of interest,
$x$. The zero of $\Dh(x,q)$ requires special treatment because
$T_L$ and $T_R$ are analytic functions except at the zero and 
the imaginary part of $T_L$ and $T_R$ are generated exclusively at the
zero. The pole of $\Dh(x,q)$ requires special treatment because the
naive expansion fails. 
To solve these problems, $T_L$ and $T_R$ need to be represented
near the transition points using comparison equations.

Near the pole of $\Dh$, the comparison
equation is $\partial_x (d/x) \partial_x T + T = 0$,
which transforms to 
$\partial_y (1/y) \partial_y T - T = 0$ under the transformation
$y=-d^{-1/3} x$. We identify the pole, $x_L$, with $x_{mn}^-(q) $
and $d \equiv A_{mn}(x_{mn}^-(q)) / \mu_{mn}'(x_{mn}^-(q))$.
Since $T_L$ increases exponentially as $x\rightarrow x_L$ from the left,
in a neighborhood of the origin, we can approximate it by 
$T_L(x) \sim T_L(x_L) Ai'(-d^{-1/3}x) /Ai'(0)$.
Using this and similar expressions, a uniformly valid asymptotic expansion
may be constructed.

\newpage
{\bf APPENDIX F: RANDOM RESONANCE BROADENING VERSUS THE 
CONTINUUM APPROXIMATION}

The continuum approximation has been justified by resonance broadening.
A specific type of resonance broadening occurs when the real eigenfrequencies,
$\omega_{m,n}$ are random, i.e. 
$\omega_{m,n}= \omega_{m,n}^o+ \delta \omega_{m,n}$. When the random
piece of the eigenfrequenies, $\delta \omega_{m,n}$, is time independent,
the resonances will be sharp for each distinct realization of 
$\delta \omega_{m,n}$, but the ensemble average will be smoother due
to blurring effects.
The additional smoothness, which an ensemble averaged solution
possesses, will result in a faster decay of the time dependent perturbation
of the ensemble averaged solution.

This random eigenfrequency resonance blurring is not equivalent to the
continuum approximation. The correct physical problem is to determine
the expectation of $T(x,\tb;\delta A_{m,n}, \delta \omega_{m,n})$.
The continuum approximation is essentially averaging or taking
the expectation of the the kernel
$K(x,\tb;\delta A_{m,n}, \delta \omega_{m,n})$ with respect to 
resonance broadening. However {\it the expectaion of Eq. (6) with a random
kernel is not equal to Eq. (6) with the expectation of the kernel.}

A second physical variant of Eq. (6) is the everywhere resonant perturbation.
In this case, the eigenfrequencies have a spatial variation such that
they are everywhere in resonance, i.e. 
$\umn(x) \equiv m \mu(x) - n\vzx - \omn(x) \equiv 0$. 
In this everywhere resonant case, Eq. (6) reduces to a random wave equation,
$T_{tt}(x,t)= \partial_x (\sum_{m,n} A_{m,n}(x) \partial_x T(x,t) )$ .
This random wave equation may easily be homogenized to yield
$T_{tt}(x,t)= \partial_x (A_h \partial_x T(x,t) )$,
where $A_h$ is the harmonic mean
of $ \sum_{m,n} A_{m,n}(x) $.
The characteristic loss time is the minor radius divided by the 
effective wave velocity: $a/\sqrt{A_h}$.
\newpage
\bigskip
\begin{center}
{\bf REFERENCES}
\end{center}
\begin{enumerate}
\item M. Avellaneda and A.J. Majda, Comm. in Math. Phys.,  
{\bf 14131}, 381 (1990).
\item  M.N. Rosenbluth, R.Z. Sagdeev, J.B. Taylor,
 and G.M. Zaslavkii, Nucl. Fusion,{\bf 6}, 297 (1966).
\item A.A. Galeev and R.Z. Sagdeev, in {\it Handbook of Plasma Physics},
Vol. 1, 677, edited by
R.N. Sudan and A.A. Galeev, (North Holland, New York, 1981).
\item A.B. Rechester, M.N. Rosenbluth, Phys. Rev. Lett. {\bf 40},  38
 (1978)  
\item J.A. Krommes, C. Oberman, and R.G. Kleva,  J. Plasma Physics
{\bf 30}, 11 (1983).
\item A. Bensoussan, J.L. Lions, and G. Papanicolaou, {\it Asymptotic
Analysis for Periodic Structures.} (North Holland, New York, 1978).
\item V. Zhikov, S. Kozlov, O.Oleinik, K. Ngoan,
Russ. Math. Surveys {\bf 34:5},  69.
\item L. Tartar, in {\it Homogenisation and Effective Moduli of
Materials and Media}, 228, J.L. Eriksen et al., editors, (Springer-Verlag,
New York 1986).
\item K.S. Riedel,  Phys. Fluids B {\bf 1}, 800 (1989).
\item K.S. Riedel,  Phys. Fluids B {\bf 2}, 2522 (1990).
\item L.E. Zakharov, K.S. Riedel, and S. Semenov, 
Letters in JETR {\bf 44},  31 (1986).
\item L.E. Zakharov and K.S. Riedel, 
Letters in JETR {\bf 44},  547, (1986).
\item A.B. Rechester, and R.B. White,
Phys. Rev. Lett. {\bf 44},  1586 (1980).
\item J.R. Cary, D.F. Escande, and A.D. Varga,
Phys. Rev. Lett. {\bf 65},  3132 (1990).
\item A.B. Rechester, M.N. Rosenbluth, and R.B. White,
Phys. Rev. Lett. {\bf 42},  1247
 (1979).  
\item H.E. Mynick and J.A. Krommes, Phys. Fl.
{\bf 30}, 11 (1983).
\item D.F. Duchs, A. Montvai, and C. Sack,
Plasma Physics and Contr. Nucl. Fus. {\bf 33}, 991
(1991).
\item R.H. Kraichnan, Phys. Fluids {\bf 8}, 575 (1965).
\item F. Olver, {\it Asymptotics and Special Functions}
(Academic Press, New York, 1974).
\item W. Wasow, {\it Linear Turning Point Theory} (Springer Verlag,
New York, 1985).
\item K.S. Riedel,  Phys. Fluids {\bf 29},  1104 (1986).
\end{enumerate}
\end{document}